\documentclass[oneside]{amsart}

\usepackage{amsmath, amssymb, amsthm, amscd, 
}

\keywords{Coverings of PDEs, B\"acklund transformations, 
Wahlquist-Estabrook structures, actions of Lie algebras on manifolds, 
the KdV equation, the Krichever-Novikov equation} 

\subjclass{37K30, 37K35, 35Q53}

\input diagrams

\allowdisplaybreaks[4]

\newarrow{TeXto}----{->}
\newcommand{\rto}{\rTeXto}

\newcommand{\rdto}{\rdTeXto}

\newcommand{\ldto}{\ldTeXto}

\newcommand{\eval}[2][\right]{\relax
  \ifx#1\right\relax \left.\fi#2#1\rvert}

\newcommand{\pd}{{\partial}}

\newcommand{\al}{{\alpha}}
\newcommand{\la}{{\lambda}}

\newcommand{\sbs}{\subset}

\newcommand{\er}{\eqref}
\newcommand{\cl}{\colon}

\newcommand{\CP}{\mathcal{P}}
\newcommand{\CE}{\mathcal{E}}
\newcommand{\CA}{\mathcal{A}}
\newcommand{\CB}{\mathcal{B}}
\newcommand{\CM}{\mathcal{M}}
\newcommand{\CN}{\mathcal{N}}

\newcommand{\ca}{\mathbf{A}}
\newcommand{\cb}{\mathbf{B}}

\newcommand{\zp}{\mathbb{Z}_+}

\newcommand{\sr}[1]{\xrightarrow{#1}}

\newcommand{\CR}{\mathcal{E}}

\newcommand{\CCD}{\mathcal{D}}
\newcommand{\CC}{\mathcal{C}}

\newcommand{\ad}{{\rm ad\,}}

\renewcommand{\sl}{\mathfrak{sl}}
\newcommand{\mg}{\mathfrak{g}}
\newcommand{\mf}{\mathfrak{f}}
\newcommand{\ma}{\mathfrak{a}}
\newcommand{\ms}{\mathfrak{s}}

\newcommand{\mL}{\mathfrak{L}}

\newcommand{\mR}{\mathfrak{R}}
\newcommand{\mK}{\mathfrak{K}}
\newcommand{\mN}{\mathfrak{N}}

\newcommand{\mh}{\mathfrak{h}}
\newcommand{\mi}{\mathfrak{i}}
\newcommand{\mn}{\mathfrak{n}}
\newcommand{\mq}{\mathfrak{q}}

\newcommand{\si}{\sigma}
\newcommand{\vf}{\varphi}

\newcommand{\sym}{\mathrm{Sym}\,}

\newcommand{\rk}{\mathrm{rank}\,}
\newcommand{\ev}{\mathrm{ev}}

\newcommand{\codim}{\mathrm{codim}\,}

\newcommand{\Com}{\mathbb{C}}

\newcommand{\id}{\mathrm{id}}

\newcommand{\nf}{\mathbf{INF}}

\newcommand{\cprime}{\/{\mathsurround=0pt$'$}}

\newtheorem{theorem}{Theorem}
\newtheorem{proposition}{Proposition}
\newtheorem{lemma}{Lemma}

\theoremstyle{definition}
\newtheorem{definition}{Definition}
\newtheorem{example}{Example}
\newtheorem{remark}{Remark}

\begin{document}


\author{Sergey Igonin}
\address{Department of Mathematics \\ 
Utrecht University \\
P.O. Box 80010 \\ 
3508 TA Utrecht \\ 
the Netherlands} 

\email{igonin@mccme.ru}

\title[Coverings and fundamental algebras for PDEs]
{Coverings and fundamental algebras \\ 
for partial differential equations}

\date{}

\begin{abstract}
Following I.~S.~Krasilshchik and A.~M.~Vinogradov \cite{nonl}, we
regard PDEs as infinite-dimensional manifolds with involutive 
distributions and consider their special morphisms called differential
coverings, which include constructions like Lax pairs and
B\"acklund transformations. We show that,
similarly to usual coverings in topology, at least for some PDEs
differential coverings are determined by actions of a sort of
fundamental group. This is not a group, but a certain
system of Lie algebras, which generalize Wahlquist-Estabrook algebras. 
From this we deduce an algebraic necessary
condition for two PDEs to be connected by a B\"acklund transformation.
We compute these infinite-dimensional
Lie algebras for several KdV type equations 
and prove 
non-existence of 
B\"acklund transformations. 

As a by-product, for some class of Lie algebras $\mg$ 
we prove that any subalgebra of $\mg$ of finite codimension
contains an ideal of $\mg$ of finite codimension.
\end{abstract}


\maketitle

\tableofcontents

\section*{Introduction} 

In this paper we study special correspondences called
(\emph{differential}) \emph{coverings} between systems of PDEs.
Roughly speaking, a covering $\CE_1\to\CE_2$
is a differential mapping from one system $\CE_1$
to another system $\CE_2$ such that
the preimage of each local solution of $\CE_2$
is a family of $\CE_1$ solutions dependent on
a finite number $m$ of parameters.

For example, 
if $v(x,t)$ is a solution of the modified KdV equation 
\begin{equation}
\label{mkdv}
	v_t=v_{xxx}-6v^2v_x
\end{equation}
then the function
\begin{equation}
\label{mt}
	u=v_x-v^2
\end{equation}
satisfies the KdV equation $u_t=u_{xxx}+6uu_x$. 
This is the famous Miura transformation, 
which determines a covering from the modified KdV equation to the KdV equation. 
For a given local solution $u(x,t)$ of the KdV equation, 
a one-parameter family of functions $v(x,t)$ is recovered 
from equations~\er{mt} and~\er{mkdv}. 
That is, we have $m=1$ for this covering. 
In general, systems $\CE_1$ and $\CE_2$ may be overdetermined, 
but must be consistent. 

More precisely, following~\cite{rb,nonl,86}, 
we regard $\CE_1,\,\CE_2$ as submanifolds in infinite jet spaces.
The (usually infinite-dimensional) submanifold
of infinite jets satisfying a system of PDEs
is called the \emph{infinite prolongation} of the system
and possesses
a canonical involutive distribution called the
\emph{Cartan distribution}.
This distribution is spanned by
the total derivative operators (regarded
as commuting vector fields on the infinite jet space)
with respect to the independent variables.
A (differential) covering $\tau\colon\CE_1\to\CE_2$ 
is a bundle of finite rank\footnote{One can consider also coverings of infinite rank~\cite{rb,nonl}, 
but we study only the case of finite rank} $m$ such that the differential $\tau_*$ 
maps the Cartan tangent subspaces of $\CE_1$ isomorphically 
onto the ones of $\CE_2$. 
Note that even local classification of coverings is highly nontrivial 
due to different possible configurations of the distributions. 

It was shown in \cite{nonl} that all kinds of
Lax pairs, zero-curvature representations,
Wahlquist-Estabrook prolongation structures,
and B\"ack\-lund transformations in soliton theory
are special types of coverings. 
In particular, a B\"acklund transformation
between two systems $\CE_1$ and $\CE_2$ is given by another
system $\CE_3$ and a pair of coverings
$\CE_1\leftarrow\CE_3\to\CE_2$. 


The name `coverings' for such bundles is used
because they include usual topological coverings
of finite-dimensional manifolds, see Example~\ref{tc} below.

Recall that for a finite-dimensional manifold $M$
its topological coverings are in one-to-one correspondence with
actions of the fundamental group $\pi_1(M)$ on (discrete)
sets. 
The main result of this paper is that at least for
some PDEs $\CE$ differential coverings are also 
determined by actions of a sort of fundamental group. 
However, this is not a group, 
but a certain system of Lie algebras that we call 
the \emph{fundamental algebras of} $\CE$.  
They are arranged in a sequence of epimorphisms 
\begin{equation}
\label{ifk}
\dots\to\mf_{k+1}\to\mf_k\to\dots\to\mf_1\to\mf_0. 
\end{equation}
Differential coverings of rank $m$ are determined 
by actions of these Lie algebras on $m$-dimensional 
manifolds $W$, that is, homomorphisms from $\mf_k$ to the algebra $D(W)$ 
of vector fields on $W$. Two coverings are isomorphic if and 
only if the corresponding actions are isomorphic. 

More precisely, the following facts hold: 
\begin{itemize} 
\item for each action $\rho\colon\mf_k\to D(W)$ we introduce 
an involutive distribution on the manifold $\CE\times W$ such that 
the trivial bundle $\CE\times W\to\CE$ becomes a covering denoted $\tau(\rho)$ 
(here $\CE$ is endowed with the fixed Cartan distribution), 
\item for any covering $\tau\colon\tilde\CE\to\CE$ we define an action 
$\rho(\tau)\colon\mf_k\to D(\tilde\CE)$ for some $k$ 
such that $\tau_*\rho(\tau)=0$,   
\item for an action $\rho_0\colon\mf_k\to D(W)$ and the covering 
$\tau=\tau(\rho_0)$, the action $\rho(\tau)$ is equal to the composition 
of the natural embedding $D(W)\subset D(\CE\times W)$ with the action $\rho_0$,   
\item a morphism of coverings $\tau_1$ and $\tau_2$ of $\CE$ 
induces a morphism of the actions $\rho(\tau_1)$ and $\rho(\tau_2)$, 
\item a covering $\tilde\CE\to\CE$ on a neighborhood of each point 
of $\tilde\CE$ is isomorphic to the covering $\tau(\rho)$ 
for some action $\rho$ of $\mf_k$ and some $k$. 
\end{itemize}

The algebra $\mf_0$ is equal to the Wahlquist-Estabrook 
prolongation algebra of $\CE$~\cite{nonl,kdv1,Prol}. 
To obtain algebras $\mf_k$ for $k\ge 1$, 
we replace the Wahlquist-Estabrook ansatz 
by jets of arbitrary order and find a canonical form of coverings 
with respect to the local gauge equivalence. 

Note that some similarity between Wahlquist-Estabrook 
algebras and the topological fundamental group was noticed in~\cite{86}. 
However, before the present paper this idea was not developed 
and did not lead to any applications.  

We prove that all finite-dimensional quotients of the fundamental algebras 
are coordinate-independent invariants of the system of PDEs. 
Namely, recall that quotients 
of the topological fundamental group $\pi_1(M)$ occur 
as automorphism groups of regular topological coverings of $M$. 
Similarly, finite-dimensional quotients of the 
fundamental algebras occur as Lie algebras of infinitesimal automorphisms 
of certain coverings of $\CE$. 

We conjecture that the fundamental algebras themselves are also 
coordinate-independent invariants and hope to prove this elsewhere
using the homological techniques of~\cite{hc_zcr,marvan,verb}. 

We formulate some conditions for a system of PDEs 
to possess fundamental algebras. We check these conditions 
and compute algebras~\er{ifk} for three PDEs: 
the KdV equation, the nonsingular Krichever-Novikov equation, 
and the linear equation
\begin{equation}
\label{iutu3}
u_t=u_{xxx}.
\end{equation}
In all three cases each $\mf_k$ is obtained from 
a single Lie algebra $\mK$ applying several times the operation 
of one-dimensional central extension. 
 
For the KdV equation we have
$\mK=\sl_2(\Com)\otimes_\Com\Com[\la]$. 

For the nonsingular Krichever-Novikov equation the algebra
$\mK$ is isomorphic to a certain subalgebra of the tensor product of
$\mathfrak{sl}_2(\Com)$ with the algebra of regular functions
on an affine elliptic curve. Note that in this case $\mf_0=0$, 
that is, the Wahlquist-Estabrook ansatz gives no nontrivial coverings. 

For equation~\er{iutu3} the algebra $\mK$ possesses a filtration by solvable ideals
$$
\mK_0\subset\mK_1\subset\dots\subset\mK_k\subset\dots\subset\mK
$$
such that the quotient $\mK/\cup_{k=0}^\infty\mK_k$ is solvable as well. 

The described methods to compute fundamental algebras 
can be applied to other evolution equations as well.

In order to develop this theory, we obtain the following results 
on Lie algebras, which may be of independent interest. 
A Lie algebra $\mg$ is said to be \emph{quasi-finite} if any subalgebra 
of $\mg$ of finite codimension contains an ideal of $\mg$ 
of finite codimension. We prove that 
\begin{itemize}
	\item a central extension of a quasi-finite algebra is quasi-finite, 
	\item for a finite-dimensional semisimple Lie algebra $\mg$ 
	and a commutative associative algebra $\CA$ the tensor 
	product $\mg\otimes\CA$ regarded as a Lie algebra is quasi-finite, 
	\item the algebra $\mK$ of the nonsingular Krichever-Novikov equation 
        is quasi-finite. 
\end{itemize}

Recall that for a connected topological covering $\tilde M\to M$ one 
has $\pi_1(\tilde M)\subset\pi_1(M)$. It turns out that 
some analog of this property is also valid for differential coverings, 
see Theorems~\ref{fpsce} and~\ref{fpsce1}. 
 
We obtain also a necessary condition for two systems of PDEs possessing 
fundamental algebras to be connected by a B\"acklund transformation: 
their fundamental algebras have to be similar in a certain sense, 
see Theorem~\ref{cbt}. As an example of using this necessary condition, 
we prove that equation~\er{iutu3} 
is not connected by any B\"acklund transformation neither 
with the KdV equation nor with the nonsingular Krichever-Novikov equation.
Note that this is apparently the first rigorous non-existence result 
for B\"acklund transformations. 
 
In this paper we consider only complex-analytic PDEs. 
Generalization of this theory to smooth PDEs is possible, 
but is a little more technical, since the analogs of Proposition~\ref{orb} 
and Theorem~\ref{gih} for smooth manifolds do not hold. 
However, practically all results will remain valid  
in the smooth case if one excludes from considered manifolds 
a thin subset of degenerate points.
Detailed exposition for smooth PDEs will be done elsewhere.

\section{Basics}
\label{psc}


In this section we review some notions of PDE geometry, actions 
of Lie algebras on manifolds and prove auxiliary lemmas 
needed for further theory.

In Subsections~\ref{inf}--\ref{ctpde} 
we mainly follow \cite{rb,nonl,kv}.
However, there are certain modifications because of the fact 
that we deal with complex-analytic manifolds, 
while in~\cite{rb,nonl,kv} only smooth manifolds 
are considered. In particular, we have to use sheaves 
instead of globally defined functions. 
Besides, the notions of subequations and irreducible equations are new. 

Most of the notions of Subsection~\ref{alam} are 
studied in more detail in~\cite{lie}.

In order to be more readable, 
all concepts of PDE geometry are introduced 
in two ways: invariant and coordinate.

\subsection{Some terminology}\label{subs:notation}

In this paper 
all manifolds, functions, vector fields, 
and mappings are supposed to be complex-analytic.

For a manifold $M$ we denote by 
$D(M)$ the Lie algebra of vector fields on $M$.
For a function $f$ on $M$ and a point $a\in M$,  
the differential of $f$ at $a$ is denoted by $d_af$. 

The differential of a mapping $\vf\colon M_1\to M_2$ 
of manifolds is denoted by $\vf_*$.

$\zp$ is the set of nonnegative integers. 

For subspaces $V_1,\dots,V_k$ of a linear space, 
the space $\langle V_1,\dots,V_k\rangle$ is 
the linear span of $V_1,\dots,V_k$. 

In this paper a surjective submersion is called a bundle. 
To emphasize its properties that in Section~\ref{inf} will 
be extended to infinite-dimensional manifolds, 
we give the following definition. 

\begin{definition}
\label{bundle}
A mapping $\varphi\colon M_1\to M_2$ of manifolds is called 
a \emph{bundle} if 
\begin{itemize}
	\item the mapping $\vf$ is surjective,
	\item for any point $a\in M_1$ 
	there is a  neighborhood $a\in U\subset M_1$ and  
	a manifold $W$ such that $\vf(U)$ is open in $M_2$ 
	and one has the commutative diagram 
	\begin{diagram}[small]
U&      &\rto^{\xi} &      &\vf(U)\times W\\
     &\rdto_{\vf}& & \ldto_p&\\
     &         &\vf(U)&      &
\end{diagram}
where $\xi$ is a complex-analytic diffeomorphism and $p$ is the projection 
to the first factor.

In this case the preimages $\vf^{-1}(b)$ of points 
$b\in M_2$ are submanifolds in $M_1$ and are 
called the \emph{fibres} of $\vf$.
They are not necessarily isomorphic to each other, 
but have the same dimension called the \emph{rank} 
of $\vf$.
\end{itemize}
\end{definition}

For a bundle $\varphi\colon M_1\to M_2$, 
a vector field $V\in D(M_1)$ is said to be \emph{$\vf$-vertical} 
if $\vf_*(V)=0$.

In what follows we say that a certain property holds \emph{locally} 
if it holds on a neighborhood 
of each point of the manifold under consideration.

\subsection{Infinite-dimensional manifolds}
\label{inf}

We want to extend the category of finite-di\-men\-si\-o\-nal 
manifolds in order to include certain type of 
infinite-dimensional manifolds that occur in PDE geometry. 

\begin{definition}
Define a category $\nf$ as follows. 
\begin{itemize}
	\item First, an \emph{elementary object} of $\nf$ is an infinite chain of bundles 
\begin{equation}
\label{obj}
	\sr{\vf_{i+2,i+1}}M^{i+1}\sr{\vf_{i+1,i}}M^{i}\sr{\vf_{i,i-1}}\dots\sr{\vf_{1,0}}M^0,
\end{equation}
where $M^i$ are  finite-dimensional manifolds. 

Two elementary objects  
$$
  \{M_1^i,\vf_{i+1,i}^1\},\quad \{M_2^i,\vf_{i+1,i}^2\}
$$
such that 
\begin{equation*}
\exists\,p,q\in\mathbb{Z}\quad M_1^{i+q}=M_2^i,\ \vf_{i+q+1,i+q}^1=\vf_{i+1,i}^2
\quad \forall\,i\ge p
\end{equation*}
are regarded to be identical. 

Denote by $\CM$ elementary object~\er{obj}. 
A \emph{point} of $\CM$ is a sequence 
\begin{equation}
\label{point}
(a_0,a_{1},\dots,a_i,\dots),\quad a_i\in M^i,\quad 
\vf_{i+1,i}(a_{i+1})=a_i\quad\forall\,i\ge 0.	
\end{equation}
Let us introduce a topology on the set $|\CM|$ of points of $\CM$.
Let $U$ be an open subset of some $M^p$. 
Denote by $U_i,\,i\ge p$, the preimage of $U$ 
in $M^{p+i}$ under bundles~\er{obj}. 
The subset of points~\er{point} such that 
$a_i\in U_i$ for all $i\ge p$
is called the \emph{elementary open subset of $|\CM|$ corresponding to $U$} 
and is denoted by $S(U)$.
By definition, elementary open subsets form a base of the topology 
on $|\CM|$. 

Let us define the structure sheaf of functions on $|\CM|$. 
Each (complex-analytic) function $f\cl U\to\Com$ 
determines the following function on $S(U)$  
\begin{equation*}
(a_0,a_{1},\dots,a_i,\dots)\mapsto f(a_p).	
\end{equation*}
Such functions on $S(U)$ are said to be \emph{elementary}.
Now let $Z$ be an open subset of $|\CM|$. A function 
$g\cl Z\to\Com$ belongs to the structure sheaf 
if and only if for each point $a\in Z$ there is 
an elementary open subset $S(U)$ such that $a\in S(U)\subset Z$ 
and the restriction of $g$ to $S(U)$ is an elementary 
function.
  \item If 
  $$
  \CM_1=\{M_1^i,\vf_{i+1,i}^1\},\quad 
  \CM_2=\{M_2^i,\vf_{i+1,i}^2\}
  $$
  are two elementary objects of $\nf$ then a \emph{morphism} $\psi\colon\CM_1\to\CM_2$ 
  is given by $\al,k\in\mathbb{Z}$ and a system of maps 
\begin{equation*}
	\psi_i\colon M_1^{i+\al}\to M_2^{i},\quad i\ge k,
\end{equation*}
satisfying 
$$
\forall\,i\ge k\quad 
\vf_{i+1,i}^2\circ\psi_{i+1}=\psi_i\circ\vf_{i+\al+1,i+\al}^1.
$$
\item Now an \emph{object} of $\nf$ is a topological space 
with a sheaf of complex-valued functions that 
is locally isomorphic to an elementary object of $\nf$. 
A mapping of objects of $\nf$ is a \emph{morphism} if locally
it is a morphism of elementary objects.
\end{itemize}
\end{definition}

\begin{remark}
Although this definition is rather sketchy, 
it is sufficient for us, because all objects of $\nf$
considered in this paper are open subobjects of 
elementary objects.
\end{remark}

\begin{example}
With each finite-dimensional manifold $M$ we associate the following 
elementary object of $\nf$
\begin{equation*}
	\to M\to M\to \dots \to M,
\end{equation*}
where all arrows are the identity mappings.
This construction identifies the category of finite-dimensional manifolds 
with a subcategory of $\nf$. 	
\end{example}

Let $\CM$ be an object of $\nf$. 
The sheaf of \emph{vector fields} on $\CM$ 
is defined in the standard way as the sheaf of 
derivations of the structure sheaf. It is a sheaf of modules 
over the structure sheaf of algebras.

In particular, if $\CM$ is elementary object~\er{obj} 
then a \emph{tangent vector} at a point~\er{point} of $\CM$ 
is a sequence 
\begin{equation*}
(v_0,v_{1},\dots,v_i,\dots),\quad v_i\in T_{a_i}M^i,\quad 
(\vf_{i+1,i})_*(v_{i+1})=v_i\quad\forall\,i\ge 0.	
\end{equation*}
The vector space of all tangent vectors at a point $a$ 
is denoted by $T_a\CM$.

A \emph{distribution} on $\CM$ is 
a locally free subsheaf of submodules of the vector fields sheaf. 
In other words, a distribution $\CCD$ of rank $k$ distinguishes 
for each point $a$ of $\CM$ a subspace 
$$
\CCD_a\subset T_a\CM,\quad\dim\CCD_a=k,
$$ 
such that locally 
there are vector fields $X_1,\dots,X_k$ 
that span the subspaces $\CCD_a$.


For a finite-dimensional manifold $W$ and 
an object $\CM$ of $\nf$, 
one defines the object $\CM\times W$ of $\nf$ as follows. 
It is sufficient to consider the case when 
$\CM$ is elementary object~\er{obj}. 
Then $\CM\times W$ is the elementary object 
\begin{equation*}
	\to M^{i+1}\times W\sr{\vf_{i+1,i}\times\id}M^{i}\times W\to\dots\to M^0\times W.
\end{equation*}
Now one easily extends Definition~\ref{bundle} of bundles to the case when 
$M_1,\,M_2$ are objects of $\nf$. However, 
we always assume the fibres $W$ to be finite-dimensional manifolds. 
 
In what follows, when we speak of functions on an object of $\nf$, 
we always assume that the functions belong to the structure sheaf. 

For the sake of simplicity, below
objects of $\nf$ are also called manifolds, 
and morphisms of $\nf$ are called mappings.


\subsection{PDEs as manifolds with dis\-tri\-bu\-tions}
\label{demd}


Let $\pi\colon E\to M$ be a bundle of finite-dimensional manifolds 
and 
$$
\theta\in E,\quad \pi(\theta)=x\in M.
$$ 
Consider
a local section $f$ of $\pi$ whose graph passes through the point $\theta$.
Denote by $[f]_x^k$ the class  of all local sections whose graphs 
are tangent to the graph of $f$ at $\theta$ with order $\geq k$. The set
\[
J^k(\pi)=\{\,[f]_x^k\mid f\text{ is a local section of }\pi,\ x\in M\,\}
\]
carries a natural structure of a manifold and is called the \emph{manifold of $k$-jets}
of the bundle $\pi$. Moreover, the natural projections
\begin{gather*}
\pi_k\colon J^k(\pi)\to M,\quad [f]_x^k\mapsto x,\\
\pi_{k,k-1}\colon J^k(\pi)\to J^{k-1}(\pi),
\quad [f]_x^k\mapsto [f]_x^{k-1},
\end{gather*}
are bundles. The infinite sequence of bundles
\begin{equation}
\label{ppp}
\dots\to J^k(\pi)\xrightarrow{\pi_{k,k-1}}
J^{k-1}(\pi)\to\dots\to J^1(\pi)\xrightarrow{\pi_{1,0}}J^0(\pi)=E	
\end{equation}
determines an object of $\nf$ that is called the \emph{manifold of infinite jets} 
of $\pi$ and is denoted by $J^\infty(\pi)$.

For each local section $f$ of $\pi$ we have the local sections 
$$
j_k (f)\cl M\to J^k(\pi),\quad x\mapsto [f]_x^k,
$$
of the bundles $\pi_k,\,k=0,1,\dots$. 
These sections determine the local section 
\begin{equation*}
j_\infty (f)\colon M\to J^\infty(\pi) 	
\end{equation*}
of the natural projection $\pi_\infty\colon J^\infty(\pi)\to M$.

There is a unique distribution 
$\CC$ on $J^\infty(\pi)$ such that for any point $x\in M$ 
and any local section $f$ of $\pi$ over a neighborhood of $x$ 
we have 
\begin{equation}
\CC_{j_\infty (f)(x)}=j_\infty(f)_*\bigl(T_x M\bigl). 	
\end{equation}
This distribution is of rank $\dim M$ and is called the \emph{Cartan distribution  
of $J^\infty(\pi)$}.



Consider a system of PDEs of order $k$ imposed on sections of the bundle $\pi$.
We assume that it determines a submanifold $\CE^0\sbs J^k(\pi)$ of the manifold $J^k(\pi)$ 
such that the mapping $\pi_k\bigl|_{\CE^0}\colon\CE^0\to M$	is a bundle.  
Then a local section $f$ of $\pi$ is a solution of the system of PDEs if and only if 
the graph of $j_k(f)$ is contained in $\CE^0$. 

For each $l\in\zp$ the \emph{$l$-th prolongation} of $\CE^0$ is the set
\begin{multline*}
\CE^l=\{\,[f]_x^{k+l}\in J^{k+l}(\pi)\,\mid\,\text{the graph of }j_k(f)
\text{ is tangent to }\CE^0\\
\text{ with order }\geq l\text{ at }[f]_x^k\in\CE^0\,\},
\end{multline*}
$l=0,1,\dots$. Restricting the maps $\pi_{k+l,k+l-1}$ to $\CE^l$
and preserving the same notation for these restrictions, we obtain the sequence of maps
\begin{equation}
\label{eee}
\dots\to\CE^l\xrightarrow{\pi_{k+l,k+l-1}}\CE^{l-1}\to\dots\to\CE^0. 	
\end{equation}
Imposing natural conditions of regularity, 
we assume that all $\CE^l$ are submanifolds of $J^{k+l}(\pi)$, 
while mappings~\er{eee} are bundles. 
The obtained object $\CE$ of $\nf$ is called the \emph{infinite prolongation} 
of the initial system of PDEs.

In what follows all considered systems of PDEs are supposed to satisfy 
these regularity assumptions and, therefore, possess infinite prolongations. 
Below such object $\CE$ of $\nf$ is sometimes simply called an \emph{equation}.

The distribution $\CC$ is tangent to $\CE$. 
Its restriction to $\CE$ is denoted by $\CC_\CE$ and 
is called the \emph{Cartan distribution of $\CE$}.
It satisfies $[\CC_\CE,\CC_\CE]\subset \CC_\CE$. 
Since $\CE$ is infinite-dimensional, this does not generally
imply existence and uniqueness of maximal integral submanifolds.




\begin{definition}
\label{subeq}
Let $\CE$ be an object of $\nf$ and $\CCD$ be a distribution on it. 
A subset $\CE'\subset\CE$ is called a \emph{subequation} 
of the pair $(\CE,\CCD)$ if $\CE'$ is a submanifold of codimension $l<\infty$ 
and $\CCD$ is tangent to $\CE'$. 
More precisely, this means the following.
We have $\CE'\neq\emptyset$, and
for each point $a\in\CE'$ there are a neighborhood $a\in U\subset\CE$ 
and functions $f_1,\dots,f_l$ on $U$ such that
\begin{itemize}
	\item $\CE'\cap U=\{q\in U\,|\,f_1(q)=\dots=f_l(q)=0\}$,
	\item for any $b\in U$ the differentials 
	$
	d_bf_1,\dots,d_bf_l\in T^*_b\CE
	$ 
	are linearly independent, 
	\item the ideal of functions on $U$ generated by $f_1,\dots,f_l$ 
	is preserved by the action of vector fields from $\CCD$.
\end{itemize}	
In this case $\CE'$ is also an object of $\nf$ with 
the distribution $\CCD\bigl|_{\CE'}$. 
The number $l$ is called the \emph{codimension} 
of the subequation $\CE'$. 

A pair $(\CE,\CCD)$ is said to be \emph{irreducible} 
if $\CE$ is connected as a topological space and there 
is no subequation $\CE'\subset\CE$ of finite nonzero codimension. 
\end{definition}

Let $\CE$ be the infinite prolongation  of a system of PDEs.
Then \emph{subequations} of $\CE$ are subequations of the pair $(\CE,\CC_\CE)$, 
and $\CE$ is called \emph{irreducible} if the pair $(\CE,\CC_\CE)$ is irreducible.

\begin{remark}
The term `subequation' is motivated by the fact that a pair $(\CE,\CC_\CE)$, 
as we agreed above, is sometimes called an equation.
\end{remark}

\subsection{Coordinate description}
\label{crd}

Consider a bundle $\pi\colon E\to M$.
Let $x_1,\dots,x_n$ be local coordinates in $M$ and
$u^1,\dots,u^d$ be local coordinates in fibres of $\pi$.
For a symmetric multi-index $\si=i_1\dots i_k$ set
\begin{equation}
\label{usi}
u^j_\si=\frac{\pd^{k} u^j}{\pd x_{i_1}\dots\pd x_{i_k}}.
\end{equation}
These functions along with $x_1,\dots,x_n$ form a system of local coordinates
for the infinite-dimensional space $J^\infty(\pi)$. 
The topology on $J^\infty(\pi)$ is the following.
Choose a finite number 
$u^{j_1}_{\si_1},\dots,u^{j_r}_{\si_r}$ 
of coordinates~\er{usi} and consider the mapping
\begin{equation*}
	J^\infty(\pi)\to\mathbb{C}^{n+r},\quad 
	a\mapsto\bigl(x_1(a),\dots,x_n(a),u^{j_1}_{\si_1}(a),\dots,u^{j_r}_{\si_r}(a)\bigl).
\end{equation*}
The preimages of open subsets of $\mathbb{C}^{n+r},\,r\in\zp$, 
under such mappings are by definition open subsets of $J^\infty(\pi)$ 
and form a base of the topology on $J^\infty(\pi)$. 
Admissible functions on open subsets of $J^\infty(\pi)$ may depend 
on $x_1,\dots,x_n$ and a finite number of coordinates~\er{usi}.
Below all functions are supposed to be admissible.

The \emph{total derivative operators}
\begin{equation}
\label{dxi}
D_{x_i}=\frac{\pd}{\pd x_i}+\sum_{\si,j}u^j_{\si i}\frac{\pd}{\pd u^j_\si},
\quad i=1,\dots,n,
\end{equation}
are commuting vector fields on $J^\infty(\pi)$ and span the Cartan distribution. 

Consider a system of PDEs 
\begin{equation}
        \label{sys}
        F_\al(x_i,u^k,u^j_\si,\dots)=0,\quad \al=1,\dots,s,
\end{equation}
in the bundle $\pi$.
The basic idea of the described approach is to treat~\er{sys} 
not as differential equations in $u^k$, but as analytic 
equations in variables~\er{usi} and $x_i$.

The \emph{differential consequences} of~\er{sys} are
\begin{equation}
\label{df0}
D_{x_{i_1}}\dots D_{x_{i_r}}(F_\al)=0,\quad  
i_k=1,\dots,n,\,\ \al=1,\dots,s,\,\ r=0,1,\dots.	
\end{equation}
The infinite prolongation $\CE\subset J^\infty(\pi)$ of system~\er{sys}
is distinguished by equations~\er{df0}.
The vector fields $D_{x_i}$ are tangent to ${\mathcal{E}}$,
and their restrictions to $\CE$ will be denoted by the same symbol $D_{x_i}$. 
They span the Cartan distribution $\CC_\CE$ of $\CE$.


\begin{example}
\label{ev_eq}
Consider a scalar evolution equation in two independent variables $x,\,t$
\begin{equation}
\label{gee}
        u_t=F(x,t,u,u_1,u_2,\dots,u_p),\quad u_k=
\frac{\partial^k u}{\partial x^k},\quad u=u_0.
\end{equation}
Its infinite prolongation has the natural coordinates $x,\,t,\,u_k,\,k\ge 0$, 
since using differential consequences of~\eqref{gee} 
all $t$-derivatives are expressed 
in terms of these. 
The total derivative operators
are written in these coordinates as follows
\begin{equation*}
  D_x=\frac{\pd}{\pd x}+\sum_{j\ge 0} u_{j+1}\frac{\pd}{\pd u_j},\quad
  D_t=\frac{\pd}{\pd t}+\sum_{j\ge 0} D_x^j(F)\frac{\pd}{\pd u_j}.
\end{equation*}
\end{example}

\subsection{Differential coverings}
\label{tdc}


\begin{definition}
Let $\CE$ be an object of $\nf$ endowed with 
a distribution $\CCD$ such that 
$[\CCD,\CCD]\subset\CCD$. 
A (\emph{differential}) \emph{covering} of (or over) the pair 
$(\CE,\CCD)$ is given by a bundle of finite rank
\begin{equation}
\label{tau}
\tau\colon\tilde{\mathcal{E}}\to{\mathcal{E}}
\end{equation}
and a distribution $\CCD^\tau$ on $\tilde\CE$ such that 
\begin{itemize}
	\item $[\CCD^\tau,\CCD^\tau]\subset\CCD^\tau$,
	\item for each $a\in\tilde\CE$ 
the differential $\tau_*$ maps the space $\bigl(\CCD^\tau\bigl)_a\subset T_a\tilde\CE$ 
isomorphically onto the space $\CCD_{\tau(a)}\subset T_{\tau(a)}\CE$. 
\end{itemize}
\end{definition}

An invertible mapping 
$\varphi\colon \tilde{\mathcal{E}}\to\tilde{\mathcal{E}}$ such
that $\tau\circ\varphi=\tau$ is called a \emph{gauge transformation}.
The covering given by the same bundle $\tau$ and 
the new distribution $\varphi_*(\CCD^\tau)$ on $\tilde\CE$ is said 
to be (\emph{gauge}) \emph{equivalent} to the initial covering.

Similarly, a \emph{morphism} between two coverings
$\tau_i\colon\CE_i\to\CE,\ i=1,2$,
over the same pair $(\CE,\CCD)$ is a mapping $\vf\colon\CE_1\to\CE_2$ such that
$\tau_1=\tau_2\circ\vf$ and $\vf_*(\CCD^{\tau_1})\subset\CCD^{\tau_2}$.

A $\tau$-vertical vector field $X\in D(\tilde\CE)$ 
is called a (\emph{gauge}) \emph{symmetry} of $\tau$ if
$[X,\CCD^\tau]\subset\CCD^\tau$.
This means that the local flow of $X$ (if it exists) 
consists of automorphisms of $\tau$.
The Lie algebra of symmetries is denoted by $\sym\tau$.

Covering~\er{tau} is said to be \emph{irreducible} 
if both pairs $(\CE,\CCD)$ and $(\tilde\CE,\CCD^\tau)$ are irreducible.

\begin{example}
\label{tc}
Let us show that usual topological coverings are a particular case
of this construction.
Let $M$ be a finite-dimensional manifold and $\CCD$ 
be the whole tangent bundle of $M$.
Coverings of rank $0$ over $(M,\CCD)$ are just topological
coverings $\tau\colon\tilde M\to M$,
where $\dim\tilde M=\dim M$ and $\CCD^\tau$ 
is the whole tangent bundle of $\tilde M$.
\end{example}

If the distribution on $\CE$ is clearly fixed, 
we speak of coverings over $\CE$ (without mentioning the distribution).


Let now $\CE$ be the infinite prolongation 
of a system of PDEs~\er{sys}.
In this case we fix $\CCD$ to be the Cartan distribution $\CC_\CE$.

Let us describe a covering~\er{tau} in local coordinates.
Recall that locally $\CC_\CE$ is spanned by $D_{x_i}$.
Therefore, locally there is a unique $n$-tuple of vector fields 
\begin{equation}
\label{tdxi}
\tilde D_{x_i}\in\CCD^\tau,\quad i=1,\dots,n,	
\end{equation}
on the manifold $\tilde{\mathcal{E}}$ such that
\begin{gather}
  \label{proj}
  \tau_*(\tilde D_{x_i})=D_{x_i},\\
  \label{comm1}
  [\tilde D_{x_i},\,\tilde D_{x_j}]=0,\quad \forall\,i,j=1,\dots,n.
\end{gather}
Moreover, vector fields~\er{tdxi} span the distribution~$\CCD^\tau$. 

If $X\in\sym\tau$ then we have
\begin{equation}
\label{symx}
[X,\tilde D_{x_i}]=0,\quad i=1,\dots,n.
\end{equation}



Below in this section 
we consider equations in two independent variables
$x$ and $t$, i.e., $n=2$. Locally the bundle $\tau$ is trivial
\begin{equation}
\label{trb}
\tau\colon{\mathcal{E}}\times W\to{\mathcal{E}},
\quad \dim W=m<\infty.
\end{equation}
Let $w^1,\dots,w^m$ be local coordinates in $W$.

From~\eqref{proj} we have
\begin{equation}
\label{tilde}
        \tilde D_x=D_x+A,\quad \tilde D_t=D_t+B,
\end{equation}
where
\begin{equation}
\label{AB}
A=\sum_{j=1}^m a^j\frac{\partial}{\pd w^j},\quad 
B=\sum_{j=1}^m b^j\frac{\partial}{\pd w^j}
\end{equation}
are $\tau$-vertical vector fields on $\CE\times W$.
Condition~\eqref{comm1} is written as
\begin{equation}
  \label{cov}
  D_x B-D_t A+[A,B]=0,
\end{equation}
where 
$$
D_x B=\sum_{j=1}^m D_x(b^j)\frac{\partial}{\pd w^j},\quad
D_t A=\sum_{j=1}^m D_t(a^j)\frac{\partial}{\pd w^j}.
$$
A covering equivalent to the one given by
$A=B=0$ is called \emph{trivial}.

The manifold ${\mathcal{E}}\times W$ is itself isomorphic to
the infinite prolongation of the system that consists
of equations~\eqref{sys} and the following additional equations
\begin{equation}
\begin{aligned}
\label{naiv}
        \frac{\partial w^j}{\partial x}&=a^j(x,t,w^k,u^i_\si,\dots),\\
        \frac{\partial w^j}{\partial t}&=b^j(x,t,w^k,u^i_\si,\dots),
\end{aligned}
\quad \quad j=1,\dots,m.
\end{equation}
This overdetermined system is consistent modulo~\eqref{sys}
if and only if \eqref{cov} holds on $\CE$.
The vector fields $D_x+A,\,D_t+B$
are the restrictions of the total derivative operators
to ${\mathcal{E}}\times W$. That is, the distribution $\CCD^\tau$ 
is the Cartan distribution of this system.

Gauge transformations correspond to invertible changes of variables
\begin{equation}
\label{gtgg}
  x\mapsto x,\ t\mapsto t,\ u^i_\si\mapsto u^i_\si,\quad
  w^j\mapsto g^j(x,t,w^k,u^i_\si,\dots),\quad j=1,\dots,m,
\end{equation}
in~\eqref{naiv}. A covering is trivial if and only if it is obtained
by such change of variables from the trivial system
$$
\frac{\partial w^j}{\partial x}=\frac{\partial w^j}{\partial t}=0,
\quad j=1,\dots,m.
$$

Therefore, 
classification of coverings over $\CE$ up to local isomorphism 
is equivalent to classification of consistent modulo~\er{sys} 
systems~\er{naiv} up to locally invertible changes of variables~\er{gtgg}. 

\begin{example}
\label{ex_gt}
Consider a covering of rank~1 
\begin{equation}
\label{ex_naiv}
        \frac{\partial w}{\partial x}=a(x,t,w,u,u_1,\dots,u_k),\quad
        \frac{\partial w}{\partial t}=b(x,t,w,u,u_1,\dots,u_k)
\end{equation}
over the infinite prolongation of equation~\er{gee}.
After a gauge transformation 	
\begin{equation*}
w\mapsto f(x,t,w,u,u_1,\dots,u_r),\quad 
\frac{\pd f}{\pd w}\neq 0,	
\end{equation*}
system~\er{ex_naiv} changes to the following system
\begin{gather*}
\begin{aligned}
        \frac{\partial w}{\partial x}&=
        \frac{1}{\frac{\pd f}{\pd w}}\Bigl(a(x,t,f,u,u_1,\dots,u_k)-D_xf\Bigl),\\
        \frac{\partial w}{\partial t}&=
        \frac{1}{\frac{\pd f}{\pd w}}\Bigl(b(x,t,f,u,u_1,\dots,u_k)-D_tf\Bigl),
\end{aligned}\\
f=f(x,t,w,u,u_1,\dots,u_r),
\end{gather*}
which represents an equivalent to~\er{ex_naiv} covering.
\end{example}


Recall that in the case of two independent variables $x,t$
a \emph{conserved current} of $\CE$ 
is a pair of functions $(f,\,g)$ on $\CE$ satisfying 	
\begin{equation}
\label{ccur}
D_tf=D_xg.
\end{equation}
Two conserved currents $(f_1,\,g_1)$ and $(f_2,\,g_2)$ 
are called \emph{equivalent} if there is a function~$h$ 
such that 
\begin{equation}
\label{ffhgg}
f_2-f_1=D_x(h),\quad g_2-g_1=D_t(h).	
\end{equation}

For a conserved current~\er{ccur} 
the pair of vector fields
$$
A=f(x,t,u^i_\si,\dots)\frac{\pd}{\pd w},\quad B=g(x,t,u^i_\si,\dots)\frac{\pd}{\pd w}
$$
satisfies~\eqref{cov} and determines a covering of rank 1. 

Equivalent conserved currents~\er{ffhgg} determine equivalent coverings.
Indeed, the corresponding gauge transformation is $w\mapsto w+h$.


\subsection{Coverings as transformations of PDEs}
\label{ctpde}

Consider two systems of PDEs  
\begin{gather}
\label{sysF}
	F_\al\Bigl(x,t,u^{1},\dots,u^{d_1},\frac{\pd^{p+q}u^{j}}{\pd x^p\pd t^q},\dots\Bigl)=0,\quad 
	\al=1,\dots,s_1,\\
\label{sysG}	
	G_\al\Bigl(x,t,v^1,\dots,v^{d_2},\frac{\pd^{p+q}v^j}{\pd x^p\pd t^q},\dots\Bigl)=0,
	\quad\al=1,\dots,s_2,  
\end{gather}
and a mapping 
\begin{equation}
\label{c_vf}
	u^{j}=\varphi^{j}\Bigl(x,t,v^1,\dots,v^{d_2},\frac{\pd^{p+q}v^l}{\pd x^p\pd t^q},\dots\Bigl),\quad 
	j=1,\dots,d_1,
\end{equation}
such that the following conditions hold.
\begin{enumerate}
	\item For each local solution $v^1(x,t),\dots,v^{d_2}(x,t)$ of system~\eqref{sysG} 
	functions~\eqref{c_vf} constitute a local solution of~\eqref{sysF}.
	\item For each local solution $u^{1}(x,t),\dots,u^{d_1}(x,t)$ 
	of~\eqref{sysF} 
	the system that consists of equations~\eqref{sysG} and~\eqref{c_vf} 
	is consistent and possesses locally a general solution 
	$$
	v^1(x,t,c_1,\dots,c_{m}),\dots,v^{d_2}(x,t,c_1,\dots,c_{m})
	$$
	dependent on a finite number of complex parameters $c_1,\dots,c_m$. 
\end{enumerate}
\begin{example}
Miura transformation~\er{mt} satisfies these conditions with $m=1$.	
\end{example}

Consider the following trivial bundles
\begin{gather*}
\pi\colon\Com^{d_1+2}\to\Com^2,\quad
(x,t,u^{1},\dots,u^{d_1})\mapsto (x,t),\\
\tilde\pi\colon\Com^{d_2+2}\to\Com^2,\quad
(x,t,v^{1},\dots,v^{d_2})\mapsto (x,t),	
\end{gather*}
and their infinite jet spaces $J^\infty(\pi)$ and 
$J^\infty(\tilde\pi)$.

Denote by $D_x,\,D_t$ and $\tilde D_x,\,\tilde D_t$ the total derivative 
operators on $J^\infty(\pi)$ and $J^\infty(\tilde\pi)$ respectively. 
One has
\begin{equation*}
\frac{\pd^{p+q}u^{j}}{\pd x^p\pd t^q}=D_x^pD_t^q(u^j),\quad 
\frac{\pd^{p+q}v^j}{\pd x^p\pd t^q}=\tilde D_x^p\tilde D_t^q(v^j).	
\end{equation*}
Formulas~\er{c_vf} suggest to consider the mapping
\begin{equation}
\label{tau_jet}
\tau\colon J^\infty(\tilde\pi)\to J^\infty(\pi)	
\end{equation}
defined as follows
\begin{equation}
\label{ex_tau}
	\tau^*(x)=x,\quad \tau^*(t)=t,\quad \tau^*(u^j)=\varphi^j,\quad 
	\tau^*\Bigl(\frac{\pd^{p+q}u^{j}}{\pd x^p\pd t^q}\Bigl)=
	\tilde D_x^p\tilde D_t^q(\varphi^j). 
\end{equation}
Then we obtain 
\begin{equation}
\label{dtd}
\tau_*(\tilde D_x)=D_x,\quad
\tau_*(\tilde D_t)=D_t.	
\end{equation}

Let $\CE\subset J^\infty(\pi)$ and $\tilde\CE\subset J^\infty(\tilde\pi)$
be the infinite prolongations of systems~\er{sysF} 
and~\eqref{sysG} respectively.
Conditions~1 and~2 above need rigorous analytical 
explanation, which we do not consider. 
Instead, following~\cite{rb,nonl}, we say that 
Conditions~1 and~2 are by definition equivalent 
to the fact that $\tau(\tilde\CE)=\CE$ and the mapping
\begin{equation}
\label{tee}
\tau\bigl|_{\tilde\CE}\colon\tilde\CE\to\CE	
\end{equation}
is a bundle of rank $m$. Then from~\er{dtd} we obtain 
that~\er{tee} is a covering. 

According to construction~\er{naiv}, every covering of a system of PDEs
is locally isomorphic to a covering of this form.


\subsection{Actions of Lie algebras on manifolds}
\label{alam}

Let $\mg$ be a Lie algebra over $\Com$.
Recall that an \emph{action} of the Lie algebra $\mg$ on a 
complex manifold $W$ is a homomorphism $\mg\to D(W)$.
For $a\in W$ let 
$\ev_a\colon D(W)\to T_aW$ 
be the evaluation mapping.
For an action $\rho\colon\mg\to D(W)$ the subalgebra 
$
\{v\in\mg\,|\,\ev_a\rho(v)=0\}
$ 
is called the \emph{isotropy subalgebra} of the point $a$.

An action $\rho$ is said to be \emph{transitive} 
if the mapping $\ev_a\rho\colon\mg\to T_aW$ is surjective for each 
$a\in W$. An action $\rho$ is called \emph{free} 
if $\ker\ev_a\rho=0$ for any $a\in W$.

A bundle $W\to W'$ is called the
\emph{quotient map} with respect to an action $\rho\colon\mg\to D(W)$ 
if all vector fields from $\rho(\mg)$ are tangent to the fibres and 
the induced action on each fibre is transitive.

A \emph{morphism} from one action
$\rho_1\colon\mg\to D(W_1)$ to another action $\rho_2\colon\mg\to D(W_2)$ is
a mapping $\psi\colon W_1\to W_2$ such that 
\begin{equation}
\label{mor_cond}
\forall\,a\in W_1\,\ \forall\,v\in\mg\quad
\psi_*\bigl(\ev_a\rho_1(v)\bigl)=\ev_{\psi(a)}\rho_2(v).	
\end{equation}
The following statement is obvious.
\begin{lemma}
\label{mtao}
Let $\psi\colon W_1\to W_2$ be a morphism of transitive 
actions $\rho_i\colon\mg\to D(W_i)$, $i=1,2$. 
Then $\psi(W_1)$ is open in $W_2$.	
\end{lemma}

Let $G$ be a connected complex Lie group associated with a 
finite-dimen\-sional Lie algebra $\mg$. For $g\in G$ set
\begin{gather*}
L_g\colon G\to G,\,\ a\mapsto ga,\quad\quad\quad
R_g\colon G\to G,\,\ a\mapsto ag.	
\end{gather*}
A vector field $X\in D(G)$ is said to be \emph{right invariant} 
if 
\begin{equation}
\label{rivf}
\forall\,g\in G\quad \bigl(R_g\bigl)_*(X)=X,	
\end{equation}
and $X$ is said to be \emph{left invariant} if 
\begin{equation}
\label{livf}
\forall\,g\in G\quad \bigl(L_g\bigl)_*(X)=X.	
\end{equation}

Denote by 
$D_\mathrm{li},\,D_\mathrm{ri}\subset D(G)$ the subalgebras 
of left invariant and right invariant vector fields respectively.
It is well known that 
\begin{equation}
\label{lrg}
D_\mathrm{li}\cong D_\mathrm{ri}\cong\mg. 	
\end{equation}
and the actions of the algebras $D_\mathrm{li},\,D_\mathrm{ri}$ 
on $G$ are free and transitive.	

By isomorphisms~\er{lrg}, 
we have the free transitive action $\si\colon\mg\to D(G)$ 
of $\mg$ on $G$ by right invariant vector fields. 
Let $H\subset G$ 
be a connected Lie subgroup and $\mh\subset\mg$ 
be the corresponding Lie subalgebra.
Consider the quotient space $G/H$ 
with the canonical projection $p\colon G\to G/H$. 

Due to equation~\er{rivf}, 
all right invariant vector fields are mapped by $p_*$
to well-defined vector fields on $G/H$.
Consider the arising transitive action 
$$
\si_\mh=p_*\circ\si\colon\mg\to D(G/H)
$$  
of $\mg$ on $G/H$. The following lemma is easy to prove.
\begin{lemma}
\label{g/h}
Let $U$ be a connected open subset of $G/H$.
Let $X\in D(U)$ commute with all vector fields from
$\si_\mh(\mg)$. Then there is $V\in D_\mathrm{li}$ 
such that $X=p_*(V)$. 

And vice versa, if
$V\in D_\mathrm{li}$ is projectable
to $G/H$ then $p_*(V)$ commutes with all vector fields
from $\si_\mh(\mg)$.
An element $V\in D_\mathrm{li}\cong\mg$ is projectable
to $G/H$ if and only if $[V,\mh]\subset\mh$.

In particular, if $U$ is a connected open subset of $G$ then 
the algebra 
$$
\{V\in D(U)\,|\,[V,\si(\mg)]=0\}
$$ 
coincides with $D_{\mathrm{li}}\cong\mg$.  
\end{lemma}

\begin{lemma}
\label{fta}	
Let $W$ be a connected finite-dimensional manifold.
Suppose that an action $\rho\colon\mg\to D(W)$ is free and transitive. 
Then the Lie algebra 
$$
\{V\in D(W)\,|\,[V,\rho(\mg)]=0\}
$$
is isomorphic to $\mg$ and acts on $W$ freely and transitively as well.
\end{lemma}
\begin{proof}
It is well known that in this case
the action $\rho$ is locally isomorphic to
the action $\si\colon\mg\to D(G)$. 
By Lemma~\ref{g/h}, we obtain that for any $a\in W$ 
there is a neighborhood $a\in U\subset W$ such that 
\begin{gather*}
\{V\in D(U)\,|\,[V,\rho(\mg)]=0\}\cong\mg,\\
\forall\,b\in U\ \,\forall\,v\in T_bW\quad
\exists!\,V\in D(U):\ \ev_b V=v,\quad [V,\rho(\mg)]=0.
\end{gather*}
Since $W$ is connected, this implies the statement
of the lemma.
\end{proof}

\begin{lemma}
\label{symcom}
Let $\mg$ be a \textup{(}possibly infinite-dimensional\textup{)} 
Lie algebra, $W_1$ and $W_2$ be connected finite-dimensional manifolds, 
and $\psi\colon W_1\to W_2$ be a morphism of transitive 
actions $\rho_i\colon\mg\to D(W_i),\,i=1,2$.  
Suppose that $\psi$ is a bundle with connected fibres 
and the algebra 
$$
\ms=\{V\in D(W_1)\,|\,\psi_*(V)=0,\,[V,\rho_1(\mg)]=0\}
$$
acts freely and transitively on each fibre of $\psi$. 
Let $\mh\subset\mg$ be the isotropy subalgebra of a point $a\in W_2$ 
with respect to the action $\rho_2$. 
Then all vector fields from $\rho_1(\mh)$ are tangent to 
the fibre $F=\psi^{-1}(a)\subset W_1$ and the image 
of the algebra $\rho_1(\mh)$ in $D(F)$ is isomorphic to $\ms$. 
\end{lemma}
\begin{proof} 
The fact that all vector fields from $\rho_1(\mh)$ are tangent to $F$ 
is obvious. Denote by $\mf$ the image of $\rho_1(\mh)$ in $D(F)$. 
The algebra $\{V\in D(F)\,|\,[V,\ms]=0\}$ includes $\mf$ 
and is, by Lemma~\ref{fta}, isomorphic to $\ms$. 
Since $\dim\mf\ge\dim F=\dim\ms$, we obtain
$$
\mf=\{V\in D(F)\,|\,[V,\ms]=0\}\cong\ms. 
$$
\end{proof}

\subsection{Zero-curvature representations}
\label{zcr}
Let $\mg$ be a Lie algebra over $\Com$.
Let $\CE$ be an open subset  
of the infinite prolongation of a system of PDEs 
in two independent variables $x,\,t$ such that 
$D_x,\,D_t$ are well defined on $\CE$. 

A pair of functions
\begin{equation}
\label{mne}
M,\,N\colon\CE\to\mg	
\end{equation}
is called a \emph{$\mg$-valued
zero-curvature representation} (ZCR in short) if
\begin{equation}
\label{ddmn}
        D_x(N)-D_t(M)+[M,N]=0.
\end{equation}
We suppose that all coefficients of the 
vector-valued functions~\er{mne} are admissible 
(i.e., belong to the structure sheaf).

Then each action $\rho\colon\mg\to D(W)$ induces
the covering structure in the bundle
$\tau\colon\CE\times W\to\CE$ 
given by
$$
\tilde D_x=D_x+\rho(M),\quad\tilde D_t=D_x+\rho(N).
$$
Equation~\er{cov} for $A=\rho(M)$ and $B=\rho(N)$ follows from~\er{ddmn}. 

For a morphism of actions $\psi\colon W_1\to W_2$ the mapping
$$
\id\times\psi\colon\CE\times W_1\to\CE\times W_2
$$ 
is a morphism of the corresponding coverings.

\begin{example}
\label{otimes}
Let $\mg$ be a finite-dimensional Lie algebra. 
Clearly, a $\mg$-valued ZCR dependent polynomially on a parameter
$\la$ can be treated as a ZCR with values in the infinite-dimensional 
Lie algebra $\mg\otimes_\mathbb{C} \mathbb{C}[\la]$.
Then by the above construction each action of $\mg\otimes_\mathbb{C} \mathbb{C}[\la]$  
determines a covering. 
\end{example}

\subsection{Translation-invariant coverings}
\label{xt}

In what follows we mainly consider \emph{trans\-la\-tion-in\-va\-ri\-ant} 
PDEs~\eqref{sys} such that  
$F_\al$ do not depend on the independent variables~$x_i$. 
In this case it is convenient to exclude 
the variables $x_i$ from the set of coordinates on $J^\infty(\pi)$ 
and $\CE$.  That is, admissible functions may depend on~\er{usi}, 
but not on $x_i$. Besides, in this case we consider total derivative 
operators~\er{dxi} without the term $\pd/\pd x_i$. 

The obtained manifold and the obtained distribution on it
are called the \emph{trans\-la\-tion-in\-va\-ri\-ant infinite prolongation} 
and the \emph{trans\-la\-tion-in\-va\-ri\-ant Cartan distribution} 
of the trans\-la\-tion-in\-va\-ri\-ant system~\er{sys} 
respectively. 
Differential coverings of the trans\-la\-tion-in\-va\-ri\-ant infinite prolongation 
are called \emph{trans\-la\-tion-in\-va\-ri\-ant coverings}.

Assume that there are two independent variables~$x,\,t$. 
Then a differential covering~\eqref{naiv} is translation-invariant 
if and only if $a^j,\,b^j$ do not depend on $x,\,t$ either. 
Making this restriction, 
we in fact do not loose any coverings, since, according to~\cite{kirn}, 
with arbitrary 
covering~\eqref{naiv} of rank~$m$ we can associate the following
translation-invariant covering of rank~$m+2$
\begin{gather*}
      \frac{\partial v^1}{\partial x}=1,\quad
      \frac{\partial v^2}{\partial x}=0,\quad
      \frac{\partial w^j}{\partial x}=a^j(v^1,v^2,w^k,u^i_\si,\dots),\\
      \frac{\partial v^1}{\partial t}=0,\quad
      \frac{\partial v^2}{\partial t}=1,\quad 
      \frac{\partial w^j}{\partial t}=b^j(v^1,v^2,w^k,u^i_\si,\dots)
\end{gather*}
(we replaced $x,\,t$ by $v^1,\,v^2$ in the right-hand side of~\er{naiv}).
The fibres of this covering have the coordinates $v^1,v^2,w^1,\dots,w^m$.


\begin{example} 
\label{xtev_eq}
Consider a translation-invariant evolution equation
\begin{equation}
\label{xtgev}
        u_t=F(u,u_1,u_2,\dots,u_p),\quad u_k=
\frac{\partial^k u}{\partial x^k},\quad u=u_0.
\end{equation}
Its translation-invariant infinite prolongation has the coordinates 
$u_k$, $k\ge 0$. 
The total derivative operators
are written in these coordinates as follows
\begin{align}
  \label{dx}
  D_x&=\sum_{j\ge 0} u_{j+1}\frac{\pd}{\pd u_j},\\
  \label{dt}
  D_t&=\sum_{j\ge 0} D_x^j(F)\frac{\pd}{\pd u_j}
\end{align}
and span the translation-invariant Cartan distribution.
\end{example}

Let us rewrite the translation-invariance condition 
in coordinate-free terms.

Recall that a \emph{connection} in a bundle $\pi\colon E\to M$ 
is given by a distribution $\CCD$ on $E$ such that 
for any $a\in E$ the mapping $\pi_*\colon \CCD_a\to T_{\pi(a)}M$ 
is an isomorphism of vector spaces.
Then for each open subset $U\subset M$ we have 
the natural linear mapping
\begin{equation*}
\nabla\colon D(U)\to D(\pi^{-1}(U))
\end{equation*}
that is uniquely defined by the following condition 
\begin{equation*}
\forall\,V\in D(U)\quad \nabla(V)\in\CCD,\quad
\pi_*(\nabla(V))=V.	
\end{equation*}
The connection is said to be \emph{flat} if 
\begin{equation*}
\forall\,V_1,\,V_2\in D(U)\quad 
\nabla([V_1,V_2])=[\nabla(V_1),\nabla(V_2)].
\end{equation*}

Consider the natural mapping 
$$
\pi_{\infty,0}\colon J^\infty(\pi)\to E
$$ 
arising from~\er{ppp}. 
Let $Z$ be an open subset of $E$.
Recall~\cite{rb,86,kv} that for any vector field 
$X\in D(Z)$ there is a unique vector field 
$S(X)\in D(\pi_{\infty,0}^{-1}(Z))$ such that
\begin{equation}
[S(X),\CC]\subset\CC,\quad (\pi_{\infty,0})_*(S(X))=X,	
\end{equation}
where $\CC$ is the Cartan distribution on $J^\infty(\pi)$.

Fix a flat connection in the bundle $\pi$.
An equation $\CE\subset J^\infty(\pi)$ is said to 
be \emph{translation-invariant} (\emph{with respect to this flat connection}) 
if for any vector field $V$ on an open subset of $M$ 
the vector field $S(\nabla(V))$ is tangent to $\CE$.

Vector fields of the form $S(\nabla(V))$ span another 
distribution $\CCD'$ of rank $\dim M$ on $J^\infty(\pi)$.
Let $a\in M$. The submanifold $\CE'=\CE\cap \pi_\infty^{-1}(a)$
is the translation-invariant infinite prolongation. 
To obtain the tran\-s\-la\-tion-invariant Cartan distribution $\CC_{\CE'}$ on it, 
one projects the Cartan distribution $\CC_\CE$ 
to $\CE'$ parallel to the distribution $\CCD'$. 
The obtained distribution $\CC_{\CE'}$ is involutive, 
but may be singular at some points of $\CE'$ 
(e.g., the points $u_i=0,\,i\ge 1$, in Example~\ref{xtev_eq}), 
and we exclude these singular points from the  
translation-invariant infinite prolongation. 
It is clear from the next example that locally the structure of the pair 
$(\CE',\CC_{\CE'})$ does not depend on $a\in M$.

\begin{example}
\label{tfc}
As in Subsection~\ref{crd}, let 
\begin{equation}
\pi\colon \Com^{d+n}\to\Com^n,\quad 
(x_1,\dots,x_n,u^1,\dots,u^d)\mapsto
(x_1,\dots,x_n).	
\end{equation}
Consider the flat connection given by $\nabla({\pd}/{\pd x_i})={\pd}/{\pd x_i}$.
It is well known that locally any flat connection is isomorphic to this one.

Since we have $S({\pd}/{\pd x_i})={\pd}/{\pd x_i}$, 
an equation $\CE\subset J^\infty(\pi)$ 
is tran\-s\-la\-tion-invariant with respect to this flat connection if and only if 
it can be given by a system~\er{sys} such that
$F_\al$ do not depend on $x_i$.   
\end{example}



\subsection{Wahlquist-Estabrook coverings}

\label{wec}

Consider a translation-in\-va\-ri\-ant evolution equation~\er{xtgev} satisfying  
$\pd F/\pd u_p\neq 0$. 
In order to describe locally all its translation-invariant coverings, one must 
solve equation~\er{cov} for 
\begin{equation}
\label{bABw}
\begin{aligned}
        A&=\sum_{j=1}^m a^j(w^1,\dots,w^m,u,\dots,u_{k})\frac{\pd}{\pd w^j},\\
        B&=\sum_{j=1}^m b^j(w^1,\dots,w^m,u,\dots,u_k)\frac{\pd}{\pd w^j},	
\end{aligned}
\end{equation}
for arbitrary $k,\,m\in\zp$. 
If $k$ is less than the order $p$ of~\er{xtgev} 
then the covering is said 
to be \emph{of Wahlquist-Estabrook type}. 

Consider the following example. 
\begin{proposition}[\cite{Prol,kdv1,nonl}]
\label{c11}
For the KdV equation 
\begin{equation}
\label{kdv11}
        u_t=u_3+u_1u,\quad u_i=\frac{\pd^i u}{\pd x^i},
\end{equation}
any Wahlquist-Estabrook covering
\begin{gather*}
        D_xB-D_tA+[A,B]=0,\\
        A=A(w^1,\dots,w^m,u,u_1,u_2),\quad
        B=B(w^1,\dots,w^m,u,u_1,u_2)
\end{gather*}
is of the form 
\begin{gather}
\label{A'11}
A=X_1+\frac13uX_2+\frac16u^2X_3,\\
\label{B'11}
B=(\frac13u_2+\frac16u^2)X_2+(\frac19u^3-\frac16u_1^2+\frac13uu_2)X_3-X_4+\\
\notag
+\frac13u[X_1,[X_1,X_2]]+\frac{1}{18}u^2[X_2,[X_1,X_2]]+\frac13u_1[X_2,X_1], 
\end{gather}
where the vector fields $X_i$ depend only on $w^1,\dots,w^m$ and 
are subject to the relations 
\begin{gather}
        \label{xrel1}
        [X_1,X_3]=[X_2,X_3]=[X_1,X_4]=[X_2,[X_2,[X_2,X_1]]]=0,\\
        \label{xrel2}
        [X_1,[X_1,[X_2,X_1]]\!=\![X_4,X_2],\ 
        [X_1,[X_2,[X_2,X_1]]]\!=\![X_1,X_2]\!+\![X_4,X_3].
\end{gather}
\end{proposition}
\begin{remark}
The KdV equation~\er{kdv11} differs from the one described in the introduction, 
but one is obtained from the other by a suitable scaling transformation 
$u\mapsto cu$ for some $c\in\Com$. 
\end{remark}

Let $\mathfrak{F}$ be the free Lie algebra generated 
by the letters $X_1,\, X_2,\, X_3,\, X_4$.
Let $\mathfrak{L}$ be the quotient of $\mathfrak{F}$ 
over relations~\er{xrel1},~\er{xrel2}. 
Then formulas~\er{A'11},~\er{B'11} determine a ZCR of~\er{kdv11} 
with values in $\mL$ such that every Wahlquist-Estabrook covering 
arises from an action of $\mL$ by the construction of Section~\ref{zcr}. 
The algebra~$\mL$ is called the \emph{Wahlquist-Estabrook prolongation algebra} 
of~\er{kdv11}. 

A similar description of Wahlquist-Estabrook coverings 
is known for many equations~\er{xtgev} (see, e.g.,~\cite{dodd,kdv1,nonl}). 

Let us describe the algebra $\mathfrak{L}$ more explicitly.  
Below for $q\in\sl_2(\Com)$ and $f(\la)\in\Com[\la]$ we write the element 
$$
q\otimes f(\la)\in\sl_2(\mathbb{C})\otimes_{\mathbb{C}}\Com[\lambda]
$$ 
simply as $qf(\la)$.   

\begin{proposition}[\cite{kdv,kdv1}]
\label{eck}
The Lie algebra $\mL$ is isomorphic to the direct
sum of the Lie algebra 
$\sl_2(\mathbb{C})\otimes_{\mathbb{C}}\Com[\lambda]$ 
and the $5$-dimensional Heisenberg algebra $H$. 
The algebra $H$ has a basis
$$
r_{-3},\ r_{-1},\ r_0,\ r_1,\ r_3
$$ 
with
the commutator table $[r_{-1},r_1]=[r_3,r_{-3}]=r_0$,
the other commutators being zero. The isomorphism is given by
\begin{equation}
\label{pakdv}
X_1\!=\!r_1\!-\!\frac12y\!+\!\frac12z\la,\,\ 
X_2\!=\!r_{-1}\!+\!z,\,\ 
X_3\!=\!r_{-3},\,\ 
X_4\!=\!r_3\!-\!\frac12y\la\!+\!\frac12z\la^2,
\end{equation}
where $h,\,y,\,z$ is a basis of $\sl_2$ with the relations
$$
[h,y]=2y,\quad [h,z]=-2z,\quad [y,z]=h.
$$
\end{proposition}

\begin{remark}
One of the main ideas of this paper is to introduce 
Lie algebras playing similar role for coverings~\er{cov},~\er{bABw} 
with arbitrary $k$. 


The set of coverings of the form~\er{bABw} 
is invariant under gauge transformations of the form 
\begin{equation}
\label{gtkp}
w^i\mapsto f^i(w^1,\dots,w^m,u,\dots,u_{k-p}). 	
\end{equation} 
In order to define these Lie algebras, 
we find for coverings~\er{cov},~\er{bABw} a canonical form 
with respect to the action of gauge transformations~\er{gtkp}. 

Since for Wahlquist-Estabrook coverings 
transformations~\er{gtkp} do not depend on $u_i,\,i\ge 0$,   
all Wahlquist-Estabrook coverings are automatically in 
the canonical form. 

Coverings~\er{cov},~\er{bABw} with arbitrary $k$ 
were also studied in~\cite{finley}. 
However, gauge transformations were not considered there. 
Because of this, the authors of~\cite{finley} had to impose  
some additional constraints on vector fields~\er{bABw}. 
\end{remark}

\section{Analogs of the fundamental group for differential coverings}
\label{gtgt}

\subsection{An instructive example}

\label{fie}

To motivate the next constructions, we present 
a description of some coverings of the KdV equation 
\begin{equation}
\label{kdv1}
        u_t=u_3+u_1u.
\end{equation}
The analogous description of all translation-invariant coverings of~\er{kdv1}
will be given in Section~\ref{sec_ckdv}.

The operators $D_x,\,D_t$ below are given by~\er{dx},~\er{dt} 
with $F=u_3+u_1u$. 
\begin{theorem}
\label{c1}
Any translation-invariant covering~\eqref{cov} of the form  
\begin{equation}
\label{depend1}
        A=A(w^1,\dots,w^m,u,u_1,u_2,u_3),\quad
        B=B(w^1,\dots,w^m,u,u_1,u_2,u_3)
\end{equation}
is locally equivalent to a covering of the form
\begin{gather}
\label{A'1}
A=X_1+\frac13uX_2+\frac16u^2X_3+f_1C,\\
\label{B'1}
B=(\frac13u_2+\frac16u^2)X_2+(\frac19u^3-\frac16u_1^2+\frac13uu_2)X_3-X_4+\\
\notag
+\frac13u[X_1,[X_1,X_2]]+\frac{1}{18}u^2[X_2,[X_1,X_2]]+\frac13u_1[X_2,X_1]+g_1C, 
\end{gather}
where the vector fields $X_i,\,C$ depend only on $w^1,\dots,w^m$ 
and satisfy
\begin{equation}
\label{crel1}	
[C,X_i]=0,\quad i=1,2,3,4,
\end{equation}
in addition to relations~\er{xrel1},~\er{xrel2}. 
Here $(f_1,\,g_1)$ is a conserved current of~\eqref{kdv1} 
\begin{gather*}
        f_1=u_1^2-\frac13u^3,\quad 
        g_1=2u_1u_3-u_2^2-u^2u_2+2uu_1^2-\frac14 u^4,\\
        D_tf_1=D_xg_1.
\end{gather*}
\end{theorem}
\begin{proof}
It is easy to obtain that $A$ does not depend on $u_2,\,u_3$ and 
is a polynomial of degree 2 in $u_1$ 
\begin{equation}
\label{A2}
A=u_1^2A_2(w^1,\dots,w^m,u)+u_1A_1(w^1,\dots,w^m,u)+A_0(w^1,\dots,w^m,u).	
\end{equation}
We want to get rid of the term $u_1A_1$
by switching to a locally gauge equivalent covering. 
Namely, consider an arbitrary point 
$u_i=a_i\in\Com,\,w^j=w^j_0\in\Com$ where vector fields~\er{depend1} 
are defined. We will find a gauge transformation 
defined on a neighborhood of this point that kills the term $u_1A_1$. 

To this end, let
$$
A_1(w^1,\dots,w^m,u)=\sum_jc^j(w^1,\dots,w^m,u)\frac{\partial}{\pd w^j}.
$$
Consider the system of ordinary differential equations
\begin{equation*}
\frac{d}{du}f^j(w^1,\dots,w^m,u)=c^j(f^1,\dots,f^m,u),\quad j=1,\dots,m,
\end{equation*}
dependent on the parameters $w^1,\dots,w^m$.
Consider its local solution on a neighborhood
of the point $u=a_0,\,w^j=w^j_0$ 
with the initial condition $f^j(w^1,\dots,w^m,a_0)=w^j$.
Then the formulas
\begin{equation}
\label{kill}
  u_k\mapsto u_k,\quad w^j\mapsto f^j(w^1,\dots,w^m,u),\quad k\ge 0,\ j=1,\dots,m,
\end{equation}
define locally a gauge transformation~$\varphi$ such that
$$
\varphi_*(D_x+A)=D_x+A',\quad \varphi_*(D_t+B)=D_t+B',
$$
where the vector field
$A'$ is of the form \eqref{A2} without the linear in $u_1$ term
(compare with Example~\ref{ex_gt}).

Now it is straightforward to show that 
the vector fields $A,\,B$ are of the form \eqref{A'1}, \eqref{B'1} with 
the relations 
\begin{gather}
        \notag
        [X_2,X_3]=[X_1,X_4]=[C,X_i]=0,\quad i=1,2,3,\\
        \notag
        [C,X_4]+\frac16[X_1,X_3]=0,\\
        \notag
        [C,X_4]+\frac13[X_1,X_3]+\frac{1}{6}[X_3,[X_1,[X_1,X_2]]]
        =\frac{1}{18}[X_2,[X_2,[X_2,X_1]]],\\
        \notag
        [X_3,[X_2,[X_1,X_2]]]=0,\quad        
        [X_1,[X_1,[X_2,X_1]]=[X_4,X_2],\\
        \label{xxxc} 
        [X_1,[X_2,[X_2,X_1]]]\!=\![X_1,X_2]\!+\![X_4,X_3].
\end{gather}
From these relations it follows that 
$[X_1,X_3]$ and $(\ad^3 X_2)(X_1)$ commute with $X_1$, $X_2$.
Now applying $\ad^2 X_2$ to~\er{xxxc} we obtain $(\ad^3 X_2)(X_1)=0$,  
which implies \er{crel1}, \er{xrel1}, \er{xrel2}. 
\end{proof}



\subsection{The definition of the fundamental algebras}
\label{sdfa}

Consider a system of PDEs in two independent variables $x,\,t$.
The results of this section are applicable to the following two 
situations.
\begin{enumerate}
	\item The manifold $\CE$ is the infinite prolongation 
	defined in Section~\ref{demd}, and $\CC_\CE$ is the Cartan distribution on it.
	\item The system of PDEs is trans\-la\-tion-invariant, 
	the manifold $\CE$ is  
	the trans\-la\-tion-invariant infinite prolongation defined in Section~\ref{xt}, 
	and $\CC_\CE$ is the trans\-la\-tion-invariant Cartan distribution. 
\end{enumerate}
However, all examples of this paper belong to the second situation. 

Without loss of generality, we can assume $\CE$ to be connected. 
Moreover, we assume that the total derivative 
operators $D_x,\,D_t$ are well defined on $\CE$.
This is not a big restriction, because most of our results are local and 
locally this is always the case.  
\begin{remark}
In fact the main Definition~\ref{pfa} below can be readily generalized 
for PDEs in any number of independent variables. However, 
since all PDEs considered in this paper are in two independent variables, 
for the sake of clarity we prefer to give this simplified version. 
\end{remark}

\begin{remark}
Below in this section we use the following notation.
For an open subset $\CE'$ of $\CE$ and a finite-dimensional manifold $W$, 
the mapping 
\begin{equation}
\label{ewe}	
\CE'\times W\to\CE'
\end{equation}
is always the projection to the first factor. 
For a function $f$ on $\CE$, its restriction to $\CE'$ is denoted 
by the same symbol $f$.  
 
According to Section~\ref{tdc}, 
a covering structure in the trivial bundle~\er{ewe} is uniquely 
determined by a pair of vector fields $A,\,B\in D(\CE'\times W)$ 
that are vertical with respect to projection~\er{ewe} and 
satisfy relation~\er{cov}. 

We have the natural embedding $D(W)\subset D(\CE'\times W)$. 
A vector field $X\in D(\CE'\times W)$ belongs to $D(W)$ if and only if 
it is vertical with respect to~\er{ewe} and its coefficients 
do not depend on coordinates of $\CE$.
\end{remark}

Inspired by Theorem~\ref{c1}, let us give the following definition.

\begin{definition}
\label{pfa}
We say that $\CE$ possesses \emph{fundamental algebras} 
if there are finite sets $\CA_k,\,\CB_k$, $k\in\zp$, 
of functions on $\CE$ satisfying the relations
\begin{equation}
\label{abin}
	\CA_k\subset\CA_{k+1},\quad \CB_k\subset\CB_{k+1}\quad\forall\,k
\end{equation}
such that for any connected open subset $\CE_1$ of $\CE$
the following conditions hold.
\begin{enumerate}
\item \label{cmain}
Let $\tau\colon\tilde\CE\to\CE_1$ be a covering of $\CE_1$. 
Then for any point $a\in\tilde\CE$ there 
are a neighborhood $a\in\tilde\CE_1\subset\tilde\CE$ 
and $k\in\mathbb{N}$ such that for $\CE_2=\tau(\tilde\CE_1)\subset\CE$ 
the covering $\tau\bigl|_{\tilde\CE_1}\colon\tilde\CE_1\to\CE_2$ 	
is isomorphic to a covering $\CE_2\times W\to\CE_2$ 
of the following \emph{canonical form}
\begin{gather}
\label{cccc}
[D_x+A,\,D_t+B]=0,\\
\label{cc}
        A=\sum_{f\in\CA_k}fM_f,\quad B=\sum_{g\in\CB_k}gN_g,\\
\label{ccccc}
M_f,\,N_g\in D(W).        
\end{gather}

        \item \label{cm}
        Any morphism $\vf\colon\CE_1\times W_1\to\CE_1\times W_2$ between
        two coverings of the form 
        \begin{gather*}
        \CE_1\times W_i\to \CE_1,\quad i=1,2,\\
        A^i=\sum_{f\in\CA_k}fM^i_f,\quad
        B^i=\sum_{g\in\CB_k}gN^i_g,\\
        [D_x+A^i,\,D_t+B^i]=0,\quad M^i_f,\,N^i_g\in D(W_i),\quad i=1,2,
\end{gather*}
        is of the form $\vf=\id\times\psi$, where
        $$
        \psi\colon W_1\to W_2,\quad\psi_*(M^1_f)=M^2_f,\,\ \psi_*(N^1_g)=N^2_g.
        $$   
\item\label{csym}
Let $X\in D(\CE_1\times W)$ be a symmetry of a covering 
$\CE_1\times W\to\CE_1$ given by vector fields 
$$
D_x+A,\,D_t+B\in D(\CE_1\times W) 
$$ 
satisfying~\er{cccc},~\er{cc},~\er{ccccc}. 
Then $X\in D(W)$ and 
\begin{equation*}
[X,M_f]=[X,N_g]=0\quad \forall\,f\in \CA_k,\ \forall\,g\in\CB_k.	
\end{equation*}
\item \label{cir}
        Consider the manifold $\CE_1\times W$ with the distribution 
        spanned by $D_x+A$, $D_t+B$ of the form~\er{cccc},~\er{cc},~\er{ccccc} 
        and let $\CE'$ be a subequation of it. 
        Then locally $\CE'$ is of the form $\CE_2\times W'$, where 
        $\CE_2$ is an open subset of $\CE_1$ and $W'$ is a submanifold 
        of $W$ such that vector fields~\er{ccccc} are tangent to $W'$. 

In particular, $\CE_1\times W$ is irreducible if and only if 
$W$ is connected and 
the Lie algebra generated by vector fields~\er{ccccc}
acts on $W$ transitively.     
\end{enumerate}

In this case fundamental algebras $\mf_k$ are defined as follows. 
Let $\mq_k$ be the free Lie algebra generated by the 
letters $M_f,\,N_g$ for $f\in\CA_k,\ g\in\CB_k$. 
Let us treat~\er{cc} as functions on $\CE$ with values in $\mq_k$.
Consider the ideal $I_k$ of $\mq_k$ generated by the elements
\begin{equation*}
\sum_{g\in\CB_k}D_x(g)(a)N_g-\sum_{f\in\CA_k}D_t(f)(a)M_f
+\sum_{f\in\CA_k,\, g\in\CB_k}f(a)g(a)[M_f,N_g],\quad 
a\in\CE,
\end{equation*}
and set $\mf_k=\mq_k/I_k$.
 
Then~\eqref{cc} becomes an $\mf_k$-valued ZCR of $\CE$. 
For an action  
\begin{equation}
\label{rf}	
\rho\colon\mf_k\to D(W)
\end{equation}
denote by $\tau(\rho)$ the covering $\CE\times W\to\CE$ 
corresponding to~\er{rf}  
by the construction of Section~\ref{zcr}. 

From~\er{abin} we have the natural epimorphism 
\begin{equation}
\label{qq}
p_k\colon\mq_{k}\to\mq_{k-1}	
\end{equation}
that maps the generators
\begin{equation}
\label{kk1}
 M_f,\quad N_g,\quad f\in\CA_{k}\setminus\CA_{k-1},\quad 
 g\in\CB_{k}\setminus\CB_{k-1},
\end{equation}
to zero. 
It is easily seen that $p_k(I_k)\subset I_{k-1}$.
Therefore, epimorphisms~\er{qq} determine the epimorphisms
\begin{equation}
\label{fk}
\dots\to\mf_{k}\to\mf_{k-1}\to\dots\to\mf_1\to\mf_0.	
\end{equation}

\end{definition}

\begin{example}
\label{fkdv1}
From Theorem~\ref{c1} for the KdV equation~\er{kdv1} we can take 
\begin{gather*}
\CA_1=\{1,\,u,\,u^2,\,u^3,\,u_1^2\},\\
\CB_1=\{u^{i_0}u_1^{i_1}u_2^{i_2}u_3^{i_3}\,|\,i_n\in\zp,\ 0\le 2i_0+3i_1+4i_2+5i_3\le 8\}.
\end{gather*}
For example, in this case we have 
\begin{gather*}
M_{u_1^2}=-3M_{u^3}=\frac12N_{u_1u_3}=-N_{u_2^2}=C,\\
M_{u}=N_{u_2}=\frac13X_2,\quad N_{uu_3}=N_{u_1u_2}=0.
\end{gather*}

The algebra $\mf_1$ is isomorphic to the quotient of the free Lie algebra generated 
by the letters $X_1,\,X_2,\,X_3,\,X_4,\,C$ over relations~\er{xrel1},~\er{xrel2}, 
\er{crel1}. Formulas~\er{A'1},~\er{B'1} determine a ZCR of~\er{kdv1} with values 
in $\mf_1$ such that each covering of the form~\er{depend1} is equivalent 
to a covering determined by an action of $\mf_1$.
The algebra $\mf_0$ is isomorphic to the algebra $\mL$ 
from Section~\ref{wec}.

For $k>3$, coverings of~\er{kdv1} of the form 
\begin{equation*}
A=A(w^1,\dots,w^m,u,u_1,\dots,u_k),\quad
B=B(w^1,\dots,w^m,u,u_1,\dots,u_k)	
\end{equation*}
are determined in a similar way by actions of 
higher algebras~$\mf_{k-2}$, which will be studied 
in Section~\ref{cfakdv}.  
\end{example}

\begin{remark}
Consider the identity covering $\CE\to\CE$.
It has canonical form~\er{cccc}, \er{cc}, \er{ccccc} 
with $M_f=N_g=0$ and $W$ equal to a point. 
From Condition~\ref{cir} we 
see that any connected open subset of 
the equation $\CE$ itself must be irreducible.
\end{remark}

\begin{remark}
\label{lk}
Consider an action~\er{rf} and let $l\ge k$. 
Consider the epimorphism $\vf\colon\mf_l\to\mf_k$ 
from~\er{fk} and the action $\rho\vf\colon\mf_l\to D(W)$. 
By the construction of epimorphisms~\er{fk}, 
we have $\tau(\rho\vf)=\tau(\rho)$.

Therefore, when we consider a finite number 
of coverings determined by actions 
$$
\rho_i\colon\mf_{k_i}\to D(W_i),\quad i=1,\dots,s,
$$ 
we can assume that all the actions 
are defined on the same algebra $\mf_k$, 
where 
$$
k=\max\{k_1,\dots,k_s\}.
$$
\end{remark}

Below in this section we suppose everywhere that 
$\CE$ possesses fundamental algebras~\er{fk} and 
$\CE_1$ is a connected open subset of $\CE$. 

\begin{theorem}
\label{dcc}
For any covering $\tau\colon\tilde\CE\to\CE_1$ each point $a\in\tilde\CE$ 
lies in a locally unique irreducible subequation $\tilde\CE_a\subset\tilde\CE$.
The image $\tau(\tilde\CE_a)$ is open in $\CE_1$, and
$\tau\bigl|_{\tilde\CE_a}$ is a covering. 	
\end{theorem}
\begin{proof}
It is sufficient to prove this statement locally. 
Then we can assume that one has 
$\tilde\CE=\CE_1\times W$ and $\tau=\tau(\rho)$ 
for some action $\rho\colon\mf_k\to D(W)$. Let 
\begin{equation*}
a=(q,z)\in\CE_1\times W,\quad q\in\CE_1,\quad z\in W.	
\end{equation*}
By Proposition~\ref{orb} below, 
locally there is a unique submanifold 
$W'\subset W$ such that $z\in W'$,
all vector fields from $\rho(\mf_k)$ are tangent to $W'$, and 
the induced action on $W'$ is transitive. 
By Condition~\ref{cir} of Definition~\ref{pfa}, 
the submanifold $\tilde\CE_a=\CE\times W'\subset\tilde\CE$ 
is the required irreducible subequation. 
\begin{proposition}[\cite{orbits}]
\label{orb}
Let $\mg$ be an arbitrary Lie algebra over $\Com$
and $\rho\colon\mg\to D(W)$ be an action of $\mg$ 
on a complex-analytic manifold $W$. 
Then for each point $z\in W$ there is submanifold 
$z\in W'\subset W$ such that 
all vector fields from $\rho(\mg)$ are tangent 
to $W'$ and the action of $\mg$ on $W'$ 
is transitive. 
The submanifold $W'$ is locally unique and 
$\dim W'=\dim \ev_z\bigl(\rho(\mg)\bigl)$.
\end{proposition}
\end{proof}

Consider a covering 
$\tau\colon\tilde\CE\to\CE_1$, where 
$\tilde\CE$ is connected. 
Condition~\ref{cmain} of Definition~\ref{pfa} 
determines locally an action of $\mf_k$ on fibres of $\tau$.
Due to Condition~\ref{cm} these local actions produce 
a well-defined global action 
$$
\rho(\tau)\colon\mf_k\to D(\tilde\CE)
$$
such that $\tau_*\circ\rho(\tau)=0$.

\begin{theorem}
\label{irr=trans}
The covering $\tau$ is irreducible 
if and only if the action $\rho(\tau)$ 
is transitive on each fibre of $\tau$.	
\end{theorem}
\begin{proof}
It is sufficient to prove this locally, 
which is done similarly to the proof of Theorem~\ref{dcc}. 	
\end{proof}


\begin{theorem}
\label{2s}
Consider two coverings $\tau_i\colon\tilde\CE^i\to\CE_1,\,i=1,2$, 
and a mapping $\varphi\colon\tilde\CE^1\to\tilde\CE^2$ such that
the diagram 
\begin{diagram}[small]
\tilde\CE^1& &\rto^{\varphi}& &\tilde\CE^2\\
     &\rdto_{\tau_1} & &\ldto_{\tau_2} &\\
     &  &\CE_1& &
\end{diagram}
is commutative.
\begin{enumerate}
	\item The mapping $\varphi$ is a morphism of coverings 
	if and only if it is a morphism of the actions $\rho(\tau_1)$ 
	and $\rho(\tau_2)$.
	\item If $\tau_1$ and $\tau_2$ are irreducible and 
	$\varphi$ is a morphism of coverings then $\vf(\tilde\CE^1)$ 
	is open in $\tilde\CE^2$.
\end{enumerate}
\end{theorem}
\begin{proof}
It is sufficient to prove both statements locally. 

(1) This follows from Condition~\ref{cm} of	Definition~\ref{pfa}.

(2) This follows from the previous statement and Lemma~\ref{mtao}.
\end{proof}

\begin{remark}
Recall that a covering of a connected 
finite-dimensional manifold $M$ is connected 
if and only if the corresponding action of $\pi_1(M)$ is transitive. 
Theorem~\ref{irr=trans} suggests that in PDE geometry 
irreducible equations play the role of `connected' objects. 
Then Theorem~\ref{dcc} is the analog for PDEs of the decomposition 
into connected components of a topological space.  	
\end{remark}


\subsection{Regular coverings and their symmetry algebras}

In the present form 
the analogy of~\er{fk} with the topological fundamental group 
is not sufficiently helpful, because canonical form~\er{cc} and 
the vector fields $M_f,\,N_g\in D(W)$ 
have no invariant (coordinate-free) meaning. 
In order to recover algebras $\mf_k$ in an invariant way, recall that 
the topological fundamental group can be expressed in terms 
of automorphism groups of coverings. 
Studying differential coverings, it is more convenient
to consider infinitesimal automorphisms, i.e., symmetries.

From Condition~\ref{csym} of Definition~\ref{pfa}, 
for each action $\rho\colon\mf_k\to D(W)$ we obtain 
\begin{equation}
\label{sr}
	\sym\tau(\rho)=\{v\in D(W)\,|\,[v,\rho(\mf_k)]=0\}.
\end{equation}

Recall that a connected topological covering $\tilde M\to M$ 
is said to be \emph{regular} if the action of 
its automorphism group on $\tilde M$ is free and transitive 
on each fibre. Similarly, we call an irreducible 
differential covering $\tau\colon\tilde\CE\to\CE$ \emph{regular} 
if the action on $\tilde\CE$ of the algebra $\sym\tau$ is free
and transitive on each fibre of $\tau$.
In particular, $\tau$ is the quotient map with respect to this action,
and $\dim\sym\tau=\rk\tau$.

\begin{theorem}
\label{rca}
A covering $\tau\colon\tilde\CE\to\CE_1$ is regular 
if and only if the action on $\tilde\CE$ of the subalgebra 
$\rho(\tau)(\mf_k)\subset D(\tilde\CE)$ 
is free and transitive on each fibre of $\tau$. 
In this case one has
$\dim\rho(\tau)(\mf_k)=\rk\tau$
and $\sym\tau\cong\rho(\tau)(\mf_k)$. 
\end{theorem}
\begin{proof}
It is sufficient to prove this locally, and 
the	local version follows from~\er{sr} and Lemma~\ref{fta}.
\end{proof}

Each ideal $\mi$ of $\mf_k$ with $\codim\mi<\infty$ determines 
a regular covering as follows. 
Consider the canonical epimorphism $\psi\colon\mf_k\to\mf_k/\mi$.
Let $\si\colon\mf_k/\mi\to D(G)$ 
be the natural action by right invariant 
vector fields on the simply connected 
Lie group $G$ whose Lie algebra is 
the finite-dimensional algebra $\mf_k/\mi$. 
For any open subset $U\subset G$ we have the transitive 
action $\si\psi\colon\mf_k\to D(U)$. 
By Theorem~\ref{rca}, the corresponding covering $\tau(\si\psi)$ 
is regular, and every regular covering is locally isomorphic 
to a covering of this form. 
By Remark~\ref{lk}, if $\mi_n\subset\mf_n,\,n\ge k$, is the preimage of 
$\mi$ under epimorphism~\er{fk} then $\mf_n/\mi_n\cong\mf_k/\mi$ 
and the corresponding regular coverings are also isomorphic. 

Let $\mi_1,\,\mi_2\subset\mf_k$ be two ideals of finite codimension.
Consider the simply connected Lie groups $G_1,\,G_2$ associated 
with the Lie algebras $\mf_k/\mi_1,\,\mf_k/\mi_2$. 
Let $U_i\subset G_i,\,i=1,2$, be connected open subsets.

Suppose that the corresponding regular coverings 
are connected by a morphism $\varphi$
\begin{diagram}[small]
\CE_1\times U_1& &\rto^{\varphi}& &\CE_1\times U_2\\
     &\rdto_{\tau_1} & &\ldto_{\tau_2} &\\
     &  &\CE_1& &
\end{diagram}
From Condition~\ref{cm} of Definition~\ref{pfa} it follows 
that $\mi_1\subset\mi_2$ and $\varphi=\id\times\psi$, where 
$\psi\colon U_1\to U_2$ is a morphism of actions of $\mf_k$.
By Theorem~\ref{rca}, we have $\sym\tau_i\cong\mf_k/\mi_i,\,i=1,2$.
The mapping 
\begin{equation*}
\vf\colon\CE_1\times U_1\to\vf(\CE_1\times U_1)=\CE_1\times\psi(U_1)	
\end{equation*}
is the quotient mapping with respect to the action of the subalgebra 
$\mi_2/\mi_1\subset\sym\tau_1$ on the manifold $\CE_1\times U_1$. 

Similarly to Theorem~\ref{2s}, 
this local description of regular coverings and morphisms 
connecting them implies the following global result.
\begin{theorem}
\label{mrc}
Consider two regular coverings $\tau_i\colon\tilde\CE^i\to\CE_1,\,i=1,2$, 
and let  
\begin{diagram}[small]
\tilde\CE^1& &\rto^{\varphi}& &\tilde\CE^2\\
     &\rdto_{\tau_1} & &\ldto_{\tau_2} &\\
     &  &\CE_1& &
\end{diagram}
be a morphism of them. Then 
there is $k\in\mathbb{N}$ and two ideals 
$\mi_1,\,\mi_2$ of $\mf_k$ of finite codimension such that 
\begin{itemize}
	\item one has 
\begin{equation}
\label{tfi}
	\sym \tau_i=\mf_k/\mi_i,\quad i=1,2,
\end{equation}
\item 
we have $\mi_1\subset\mi_2$, the subset $\vf(\tilde\CE^1)$ 
is open in $\tilde\CE^2$, and the mapping
$
\vf\colon\tilde\CE^1\to\vf(\tilde\CE^1)
$
is the quotient mapping with respect to the action of the subalgebra 
$\mi_2/\mi_1\subset\sym\tau_1$ on the manifold $\tilde\CE^1$, 
\item
the differential $\vf_*$ of $\vf$ induces an epimorphism of algebras 
$\sym\tau_1\to\sym\tau_2$. In terms of isomorphisms~\er{tfi} 
it is the natural epimorphism $\mf_k/\mi_1\to\mf_k/\mi_2$ 
corresponding to the inclusion $\mi_1\subset\mi_2$. 
\end{itemize}
\end{theorem}

In contrast to fundamental algebras~\er{fk}, 
the system of symmetry algebras of regular coverings 
is a coordinate-free canonical invariant of a system of PDEs, 
since symmetry algebras are coordinate-independent objects.
Thus we recover in an invariant way not algebras~\er{fk} 
themselves, but all their finite-dimensional quotients.

\subsection{Quasi-finite Lie algebras}

We present here some results on Lie algebras.

\begin{definition}
A Lie algebra $\mg$ is said to be \emph{quasi-finite} if for 
any subalgebra $\mh\subset\mg$ of finite codimension 
there is an ideal of $\mg$ that is of finite codimension 
and is contained in $\mh$. 	
\end{definition}

\begin{theorem}
\label{gih}
Let $\mg$ be a quasi-finite algebra. 
Then for any transitive action $\rho\colon\mg\to D(W)$ 
on a connected finite-dimensional manifold $W$
the algebra $\rho(\mg)$ is finite-dimensional.
\end{theorem}
\begin{proof}
Let $a\in W$ and 
$\mh\subset\mg$ be the isotropy subalgebra of $a$.
Since $\mg$ is quasi-finite and $\codim\mh=\dim W<\infty$, there is an ideal $\mi$ 
of $\mg$ such that $\mi\subset\mh$ and $\codim\mi<\infty$. 
It is well known that in the complex-analytic situation 
the image $\rho(\mh)$ of the isotropy subalgebra cannot contain 
any nontrivial ideal of $\rho(\mg)$. Therefore, $\rho(\mi)=0$ and 
$$
\dim\rho(\mg)\le\codim\mi<\infty.
$$ 
\end{proof}

\begin{theorem}
\label{qfce}
Let $\mg$ be a quasi-finite Lie algebra 
and $\rho\colon\tilde\mg\to\mg$ 
be an epimorphism such that 
\begin{equation}
\label{kerg}
[\ker\rho,\tilde\mg]=0
\end{equation}
\textup{(}that is, $\tilde\mg$ is a central extension of $\mg$\textup{)}. 
Then $\tilde\mg$ is also quasi-finite.
\end{theorem}
\begin{proof}
Let $\mh\subset\tilde\mg$ be a subalgebra of finite codimension. 
Clearly, the subset  
\begin{equation}
\label{mh1}
\mh_1=\{a\in\mh\,|\,[\tilde\mg,a]\subset\mh]\}	
\end{equation}
is also a subalgebra of finite codimension. 
By assumption, there is an ideal $\mi$ of $\mg$ that is of finite codimension 
and is contained in $\rho(\mh_1)$. 

The subspace $\rho^{-1}(\mi)\cap\mh$ is of finite codimension 
and is contained in $\mh$. Let us prove that $\rho^{-1}(\mi)\cap\mh$ 
is an ideal of $\tilde\mg$.

Let $a\in\rho^{-1}(\mi)\cap\mh$ and $v\in\tilde\mg$. 
Then $a=h+z$, where $h\in\mh_1$ and $z\in\ker\rho$.
Combining~\er{kerg} and~\er{mh1}, 
we obtain $[v,a]\in\mh$. Besides, since $\rho^{-1}(\mi)$ is 
an ideal of $\tilde\mg$, we have $[v,a]\in\rho^{-1}(\mi)$. 
Therefore, $[v,a]\in \rho^{-1}(\mi)\cap\mh$. 
\end{proof}


Let $\mg$ be a Lie algebra over $\Com$ 
and $\CA$ be a commutative associative algebra over $\Com$. 
Then the space $\mg\otimes_{\mathbb{C}}\CA$ has the following
natural Lie algebra structure 
\begin{equation}
\label{ga}
[g_1\otimes a_1,\,g_2\otimes a_2]=[g_1,g_2]\otimes a_1a_2,
\quad g_1,\,g_2\in\mg,\ a_1,\,a_2\in\CA.	
\end{equation}

\begin{theorem}
\label{slih}
If $\mg$ is finite-dimensional and semisimple then 
the Lie algebra $\mg\otimes_{\mathbb{C}}\CA$ is quasi-finite.
\end{theorem}
\begin{proof}
Let $\mh\subset\mg\otimes_{\mathbb{C}}\CA$
be a subalgebra of finite codimension.
Then the subspace
\begin{equation}
\label{z}
Z=\{f\in\CA\,|\ \mg\otimes f\subset\mh\}
\end{equation}
is of finite codimension in $\CA$.
Since $[\mg,\mg]=\mg$, the subspace $Z$ is a
subring of $\CA$.
The subspace $Z'=\{f\in Z\,|\,f\CA\subset Z\}$ is
of finite codimension and is an ideal of the ring $\CA$.
Therefore, the subspace $\mg\otimes Z'$ is an ideal
of $\mg\otimes\CA$ of finite codimension, and 
from~\er{z} we have $\mg\otimes Z'\subset\mh$.
\end{proof}


\subsection{Local structure of irreducible coverings}

Below we suppose that algebras~\er{fk} are quasi-finite.

Consider a subalgebra $\mh\subset\mf_k$ of finite codimension.
Let $\mi(\mh)$ be the maximal ideal of $\mf_k$ that is 
contained in $\mh$. Since $\mf_k$ is quasi-finite, we have 
$\codim\mi(\mh)<\infty$.

Let $G$ be the simply connected Lie group whose Lie algebra
is $\mf_k/\mi(\mh)$ and $H\subset G$ be the connected Lie subgroup whose 
Lie subalgebra is $\mh/\mi(\mh)$.
According to Section~\ref{alam}, 
the algebra $\mf_k$ acts on $G$ by right invariant vector fields, 
which are projected also to the quotient space $G/H$. 
Denote by $\si(\mf_k,\mh)$ the arising transitive action 
of $\mf_k$ on the manifold $W(\mf_k,\mh)=G/H$. 
We have 
$\ker\si(\mf_k,\mh)=\mi(\mh)$.	
\begin{remark}
Let $G$ be a Lie group associated with a Lie algebra $\mg$. 
Generally, not for every subalgebra $\mh\subset\mg$ there is a Lie subgroup 
whose Lie subalgebra is $\mh$. However, for us it is sufficient 
to consider the local Lie subgroup, which always exists. 
In this case the symbol $G/H$ denotes the quotient space not of the whole 
group $G$, but of some neighborhood of the unity element. 	
\end{remark}

As above, let $\CE_1$ be a connected open subset of $\CE$.
Consider the manifold 
$\CE_1(\mf_k,\mh)=\CE_1\times W(\mf_k,\mh)$ and the covering 
$$
\tau(\mf_k,\mh)
\colon\CE_1(\mf_k,\mh)\to\CE_1
$$ 
corresponding to the action $\si(\mf_k,\mh)$ of $\mf_k$.
\begin{theorem}
\label{efh}
The following statements hold.
\begin{enumerate}
\item \label{tt}
Every irreducible covering $\tau$ of $\CE_1$ is locally isomorphic 
to a covering $\tau(\mf_k,\mh)$ for some $k\in\mathbb{N}$ 
and $\mh\subset\mf_k$.
\item We have 
\begin{equation}
\label{sfh}
\sym\tau(\mf_k,\mh)\cong\mn(\mh)/\mh,	
\end{equation}
where
\begin{equation}
\label{nh}
\mn(\mh)=\{v\in\mf_k\,|\,[v,\mh]\subset \mh\}.
\end{equation}
\item \label{hi}
The covering $\tau(\mf_k,\mh)$ is regular 
if and only if $\mh$ is an ideal of $\mf_k$.
\end{enumerate}
\end{theorem}
\begin{proof}

(1) By Conditions~\ref{cmain} and~\ref{cir} of Definition~\ref{pfa}, 
locally we have $\tau=\tau(\rho)$ for some transitive 
action $\rho\colon\mf_k\to D(W)$. 
Let $a\in W$ and consider the isotropy subalgebra 
$\mh\subset\mf_k$ of $a$. By Theorem~\ref{gih}, 
the actions $\rho$ and $\si(\mf_k,\mh)$ are locally isomorphic. 
Then the coverings $\tau(\rho)$ and $\tau(\mf_k,\mh)$ 
are locally isomorphic as well.
 
(2) Formulas~\er{sfh} and~\er{nh} follow from
formula~\er{sr} and Lemma~\ref{g/h}.  	

(3) This follows from Theorem~\ref{rca}.
\end{proof}

Recall that for any connected 
topological covering $\tau\colon\tilde M\to M$ there is 
a commutative diagram of coverings 
\begin{diagram}[small,LaTeXeqno]
\label{mmm}
M'&     &\rto^{p} &     &\tilde M\\
     &\rdto_{r}& &\ldto_\tau&\\
     &       &M&         &
\end{diagram}	
where $r$ is regular and $p$ is the quotient
mapping with respect to the action of some automorphism 
subgroup of $r$.
Let us construct an analog of diagram~\er{mmm} 
for differential coverings.

\begin{theorem}
\label{prsys}
Any irreducible covering $\tau\colon\tilde\CE\to\CE_1$ 
is locally included in a commutative diagram of irreducible coverings
\begin{diagram}[small]
\CR'&              &\rto^{p} &     &\tilde\CE\\
     &\rdto_{r}& &\ldto_\tau&\\
     &        &\CE_1&          &\\
\end{diagram}
such that the following assertions hold.
\begin{enumerate}
        \item The covering $r$ is regular.
        \item The covering $p$ is the quotient morphism with respect to the
        action on $\CR'$ of some subalgebra $\mh$ of $\sym r$.
        \item
        The algebra $\sym\tau$ coincides with the quotient 
        $\mn/\mh$, where
        $$
        \mn=\{v\in\sym r\,|\,[v,\mh]\subset \mh\},
        $$
        and the action of $\mn/\mh$ on $\tilde\CE$ is induced
        by the action of $\mn$ on $\CR'$.
\end{enumerate}
\end{theorem}
\begin{proof}
By Theorem~\ref{efh}~\er{tt}, it is sufficient to prove the statements
for $\tau=\tau(\mf_k,\mh_1)$, where $\mh_1$ is a subalgebra of 
$\mf_k$ of finite codimension. 

Recall that $\mi(\mh_1)$ is the maximal ideal of $\mf_k$ that 
is contained in $\mh_1$.
By Theorem~\ref{efh}~\er{hi}, the covering $r=\tau(\mf_k,\mi(\mh_1))$ 
is regular. The inclusion of Lie algebras 
$$
\mi(\mh_1)\subset\mh_1\subset\mf_k
$$
determines a surjective morphism 
$$
W(\mf_k,\mi(\mh_1))\to W(\mf_k,\mh_1)
$$ 
of actions of $\mf_k$, which determines the surjective morphism 
$$
p\colon\CE_1(\mf_k,\mi(\mh_1))\to\CE_1(\mf_k,\mh_1),\quad 
\tau(\mf_k,\mh_1)\circ p=\tau(\mf_k,\mi(\mh_1)),
$$
of the corresponding coverings.

By formulas~\er{sfh} and~\er{nh}, one has $\sym r=\mf_k/\mi(\mh_1)$.
By construction, the morphism $p$ is the quotient map with respect to 
the action of $\mh=\mh_1/\mi(\mh_1)\subset\sym r$ on the manifold
$\CE_1(\mf_k,\mi(\mh_1))$. Finally, the last statement of the 
theorem follows from formulas~\er{sfh},~\er{nh}.
\end{proof}




For a subalgebra $\mh$ of $\mf_k$ of finite codimension, 
denote by $\mh_l$ the preimage of $\mh$ in $\mf_{l},\,l\ge k$, 
under epimorphisms~\er{fk}.  
By Remark~\ref{lk}, one obtains 
$$
\CE_1(\mf_l,\mh_l)\cong\CE_1(\mf_k,\mh),\quad\tau(\mf_l,\mh_l)\cong\tau(\mf_k,\mh)
\quad\forall\,l\ge k.
$$ 

If $\tilde\mh\subset\mh_l$ is a subalgebra of finite codimension 
then we have the natural surjective morphism 
$W(\mf_l,\tilde\mh)\to W(\mf_l,\mh_l)$ of actions of $\mf_l$,  
which determines a covering 
$$
\tau(\mh,\tilde\mh)\colon\CE_1(\mf_l,\tilde\mh)\to\CE_1(\mf_l,\mh_l)
\cong\CE_1(\mf_k,\mh).
$$

Let $\mi$ be an ideal of $\mh_l$ with $\codim\mi<\infty$ 
(but not necessarily an ideal of $\mf_l$). 
By formulas~\er{sfh},~\er{nh}, the covering 
\begin{equation}
\label{thh}
\tau(\mh,\mi)\colon\CE_1(\mf_l,\mi)\to\CE_1(\mf_k,\mh)
\end{equation}
is regular and $\sym\tau(\mh,\mi)\cong\mh_l/\mi$.
The following theorem shows that locally 
every regular covering of $\CE_1(\mf_k,\mh)$ 
is of this form.


\begin{theorem}
\label{fpsce}
Consider an irreducible covering $\tau\colon\tilde\CE\to\CE_1$
and the corresponding action $\rho(\tau)\colon\mf_k\to D(\tilde\CE)$.
Let $a\in\tilde\CE$ and $\mh\subset\mf_k$ be 
the isotropy subalgebra of $a$. 

Then for any connected neighborhood $\tilde\CE'\subset\tilde\CE$ 
of $a$ the symmetry algebra of any regular covering over $\tilde\CE'$ 
is isomorphic to a finite-dimensional quotient of $\mh_l$ for some $l\ge k$.

And vice versa, for any $l\ge k$ and any ideal $\mi$ of $\mh_l$ 
of finite codimension there is a regular covering~$\tau'$ 
over a neighborhood of $a$ such that $\sym\tau'=\mh_l/\mi$.
\end{theorem}
\begin{proof}
Let $\tau'\colon\CE''\to\tilde\CE'$ be a regular covering.  
Consider the connected open subset $\CE_2=\tau(\tilde\CE')$ of $\CE_1$ 
and the commutative diagram of coverings 
\begin{diagram}[small]
\CE''&              &\rto^{\tau'} &     &\tilde\CE'\\
     &\rdto_{\tau\tau'}& &\ldto_\tau&\\
     &        &\CE_2&          &\\
\end{diagram}
Since the question is essentially local, we can assume 
that the above diagram is of the form 
\begin{diagram}[small]
\CE_2\times W_1& &\rto^{\id\times\psi}& &\CE_2\times W_2\\
     &\rdto_{\tau(\rho_1)}& &\ldto_{\tau(\rho_2)}&\\
     &        &\CE_2&          &\\
\end{diagram}
where $\rho_i\colon\mf_l\to D(W_i)$, $i=1,2$, 
are transitive actions for some $l\ge k$ and $\psi\colon W_1\to W_2$ 
is a morphism of actions. The point $a$ is of the form $a=(q,z)$, 
$q\in\CE_2$, $z\in W_2$. The algebra $h_l$ is the isotropy algebra 
of $z$ with respect to the action $\rho_2$. Then the first statement 
of the theorem follows from Lemma~\ref{symcom} for $\mg=\mf_l$. 

The second statement of the theorem follows from construction~\er{thh}. 
\end{proof}


This theorem is the analog of the fact that 
for a connected topological covering $\tilde M\to M$ one has 
$\pi_1(\tilde M)\subset\pi_1(M)$. 

Since $\mg$ in Lemma~\ref{symcom} is allowed to be infinite-dimensional, 
the first statement of Theorem~\ref{fpsce} holds 
even if algebras~\er{fk} are not quasi-finite. 
\begin{theorem}
\label{fpsce1}
In the notation of Theorem~\ref{fpsce}, 
the symmetry algebra of any regular covering over $\tilde\CE'$ 
is isomorphic to a finite-dimensional quotient of $\mh_l$ for some $l\ge k$ 
even if the fundamental algebras are not quasi-finite. 
\end{theorem}

\subsection{Necessary conditions for existence of B\"acklund transformations}
\label{cr_bt}

Consider two systems of PDEs  
\begin{equation}
\label{fi}
	\CE_i=\Bigl\{F^i_\al\Bigl(x,t,u^{i1},\dots,u^{i{d_i}},
	\frac{\pd^{p+q}u^{ij}}{\pd x^p\pd t^q},\dots\Bigl)=0,\ 
	\al=1,\dots,s_i\Bigl\},\quad i=1,2,  
\end{equation}
A \emph{B\"acklund transformation} between $\CE_1$ and $\CE_2$
is given by another system
\begin{equation}
\label{as}
	F_\al\Bigl(x,t,v^1,\dots,v^d,\frac{\pd^{p+q}v^j}{\pd x^p\pd t^q},\dots\Bigl)=0,\quad\al=1,\dots,s.
\end{equation}
and two mappings 
\begin{equation}
\label{vf}
	u^{ij}=\varphi^{ij}\Bigl(x,t,v^1,\dots,v^d,\frac{\pd^{p+q}v^l}{\pd x^p\pd t^q},\dots\Bigl),\quad 
	j=1,\dots,d_i,\ i=1,2,
\end{equation}
such that for each $i=1,2$ one has 
\begin{itemize}
	\item for each local solution $v^1(x,t),\dots,v^d(x,t)$ of~\eqref{as} functions~\eqref{vf} 
	form a local solution of~\eqref{fi},
	\item for each local solution $u^{i1}(x,t),\dots,u^{id_i}(x,t)$ of~\eqref{fi} 
	the system that consists of equations~\eqref{as} and~\eqref{vf} is consistent and 
	possesses locally a general solution 
	$$
	v^1(x,t,c_1,\dots,c_{r_i}),\dots,v^d(x,t,c_1,\dots,c_{r_i})
	$$
	dependent on a finite number of complex parameters $c_1,\dots,c_{r_i}$. 
\end{itemize}
Similarly to Section~\ref{ctpde},
these conditions mean by definition that 
the infinite prolongation $\tilde\CE$ of~\eqref{as} covers both 
$\CE_1$ and $\CE_2$ 
\begin{diagram}[small,LaTeXeqno]
\label{btrm}
        &              &\tilde\CE&              &\\
    &\ldto_{\tau_1}& &\rdto_{\tau_2}&\\
   \CE_1&             &  &              &\CE_2
\end{diagram}
where the covering $\tau_i$ is of rank $r_i$.
We allow $\CE_i$ to be not the whole infinite prolongation, 
but some nonempty open subset of it. 

If systems~\er{fi} are translation-invariant then 
we can make~\eqref{as} tran\-s\-la\-tion-invariant as well 
using the trick from Section~\ref{xt}\,: 
replace $x,\,t$ in $F_\al$ and $\vf^{ij}$ by the new dependent variables 
$w^1,\,w^2$ respectively and add to~\eqref{as} the following equations 
\begin{equation*}
	\frac{\pd w^1}{\pd x}=\frac{\pd w^2}{\pd t}=1,\quad
	\frac{\pd w^1}{\pd t}=\frac{\pd w^2}{\pd x}=0.
\end{equation*}
After this substitution coverings~\eqref{btrm} become translation-invariant.  

\begin{example}
Consider two different coverings from the modified KdV 
equation to the KdV equation 
\begin{diagram}
   &              & v_t\!=\!v_{xxx}\!-\!6v^2v_x &              &       \\
     &\ldto^{u\!=\!v_x\!-\!v^2}& &\rdto^{u\!=\!-v_x\!-\!v^2}&          \\
  u_t\!=\!u_{xxx}\!+\!6uu_x &             &  &      & u_t\!=\!u_{xxx}\!+\!6uu_x
\end{diagram}	
This diagram presents a B\"acklund auto-transformation of the KdV equation. 
See, e.g,~\cite{backl_new,backlund} for more examples of B\"acklund transformations. 
\end{example}

\begin{theorem}
\label{cbt}
Suppose that two systems $\CE_i,\,i=1,2$, possess fundamental algebras 
\begin{equation}
\label{fk2}
\dots\to\mf^i_{k+1}\to\mf^i_k\to\dots\to\mf^i_1\to\mf^i_0,\quad i=1,2,
\end{equation}
and the algebras~$\mf^1_k$ are quasi-finite. 
Let $\mg$ be a finite-dimensional Lie algebra. 
Suppose that for any $k_1,\,k_2\in\zp$ and any 
subalgebras
\begin{equation}
\label{hf2}
\mh^i\subset\mf^i_{k_i},\quad\codim\mh^i<\infty,\quad i=1,2,	
\end{equation}
there is an epimorphism $\mh^1\to\mg$, but there is no epimorphism $\mh^2\to\mg$. 
Then there is no B\"acklund transformation between $\CE_1$ and $\CE_2$. 
\end{theorem}
\begin{proof}
Suppose that there is a B\"acklund transformation.  
By the above construction, it determines a diagram~\eqref{btrm} of coverings. 
Let $a\in\tilde\CE$. By Theorem~\ref{dcc}, 
locally there is a unique irreducible subequation 
$\tilde\CE_a\subset\tilde\CE$ that contains $a$.
The subset $\tau_i(\tilde\CE_a)$ is open in $\CE_i$, 
and the coverings 
\begin{equation*}
\tau_i\bigl|_{\tilde\CE_a}\colon\tilde\CE_a\to\tau_i(\tilde\CE_a),\quad i=1,2, 
\end{equation*}
are irreducible. 
Consider the action
\begin{equation*}
\rho\bigl(\tau_1\bigl|_{\tilde\CE_a}\bigl)\colon \mf^1_{k}\to D(\tilde\CE_a) 	
\end{equation*}
and let $\mh^1\subset\mf^1_{k}$ be the isotropy subalgebra of $a$. 
By Theorem~\ref{fpsce}, an epimorphism $\mh^1\to\mg$ implies that 
over a connected neighborhood of $a\in\tilde\CE_a$ 
there is a regular covering with symmetry algebra equal to $\mg$. 
Applying Theorem~\ref{fpsce1} to this regular covering and the covering 
$\tau_2\bigl|_{\tilde\CE_a}$, we obtain that $\mg$ is isomorphic to a quotient 
of some subalgebra $\mh^2$ of $\mf^2_{l}$ of finite codimension. 
Thus we get a contradiction. 
\end{proof}




\section{Coverings of scalar evolution equations}
\label{csev}

In this section we prove some technical results, 
which will be needed in Sections~\ref{sec_ckdv} and~\ref{sec_ckn}.
Consider a translation-invariant evolution equation  
\begin{gather}
	\label{gev}
	u_t=F(u,u_1,u_2,\dots,u_p),\quad \frac{\pd F}{\pd u_p}\neq 0,\\
	\notag
	u_i=\frac{\pd^i u}{\pd x^i},\quad u=u_0.	
\end{gather}
Let $\CE$ be a connected open subset of the translation-invariant infinite prolongation 
of this equation described in Example~\ref{xtev_eq}. 


Let $W$ be a connected open subset of $\Com^m$ 
with coordinates $w^1,\dots,w^m$ and 
\begin{equation}
\label{ua}
u_i=a_i\in\Com,\quad i=0,1,\dots,	
\end{equation}
be a point of $\CE$. 
Consider a covering $\CE\times W\to\CE$ given by vector fields 	
\begin{gather}
\notag
\begin{aligned}
        A&=\sum_{j=1}^m a^j(w^1,\dots,w^m,u,\dots,u_{k})\frac{\pd}{\pd w^j},\\
        B&=\sum_{j=1}^m b^j(w^1,\dots,w^m,u,\dots,u_k)\frac{\pd}{\pd w^j},	
\end{aligned}\\
\label{gc}
        D_xB-D_tA+[A,B]=0.
\end{gather}
Below we sometimes omit the dependence on the coordinates $w^i$ 
in vector fields on $\CE\times W$.

\begin{remark}
Below in this section we say that \emph{locally there is a gauge transformation} 
with certain properties if for any $w\in W$ a gauge transformation with these properties
exists on a neighborhood of the point $(a,w)\in\CE\times W$, 
where $a$ is the fixed point~\er{ua} of $\CE$.
\end{remark}

\begin{lemma}
\label{gcf}
We have 
\begin{equation}
\label{obv}
	\frac{\pd A}{\pd u_s}=0\quad\forall\,s>k-p+1.
\end{equation}
Moreover, locally there is a gauge transformation 
$$
w^i\mapsto f^i(w^1,\dots,w^m,u,\dots,u_{k-p}),
\quad i=1,\dots,m,
$$ 
such that the transformed vector field $D_x+A$ satisfies 
for all $s\ge 1$
\begin{equation}
\label{d=0}
\frac{\pd A}{\pd u_s}(u,\dots,u_{s-1},a_s,a_{s+1},\dots,a_k)=0
\quad\forall\,u,\dots,u_{s-1}.
\end{equation}
\end{lemma}
\begin{proof}
Differentiating equation~\eqref{gc} with respect to $u_s$ for $s>k$ and 
using the form~\eqref{dx},~\eqref{dt} 
of $D_x$ and $D_t$, one immediately obtains~\eqref{obv}. 

Now suppose that~\er{d=0} holds for all $s>n$, where $0<n\le k-p+1$. 
It easily seen that this property is preserved by 
any gauge transformation of the form 
\begin{equation}
\label{gtn}
w^i\mapsto f^i(w^1,\dots,w^m,u,\dots,u_{n-1})	
\end{equation}

By induction on $k-n$, 
it remains to find a gauge transformation~\er{gtn} 
such that the transformed vector field $D_x+A$ satisfies~\er{d=0} 
for $s=n$. Let 
\begin{equation*}
\frac{\pd A}{\pd u_n}(u,\dots,u_{n-1},a_n,a_{n+1},\dots,a_k)
=\sum_jc^j(w^1,\dots,w^m,u,\dots,u_{n-1})\frac{\partial}{\pd w^j}.	
\end{equation*}
Similarly to the proof of Theorem~\ref{c1},
consider the system of ordinary differential equations
\begin{gather*}
\frac{d}{du_{n-1}}f^j(w^1,\dots,w^m,u,\dots,u_{n-1})=
c^j(f^1,\dots,f^m,u,\dots,u_{n-1}),\\
j=1,\dots,m,	
\end{gather*}
dependent on the parameters $w^1,\dots,w^m$ and $u,\dots,u_{n-2}$. 
Its local solution with the initial condition 
$$
f^j(w^1,\dots,w^m,u,\dots,u_{n-2},a_{n-1})=w^j
$$
determines the required transformation~\er{gtn}.
\end{proof}

\begin{lemma}
\label{lcm}
Consider two coverings
\begin{gather*}
        D_xB_i-D_tA_i+[A_i,B_i]=0,\\
      A_i=\sum_{j=1}^{m_i} a_i^j(w_i^1,\dots,w_i^{m_i},u,\dots,u_{k_i})\frac{\pd}{\pd w_i^j},\\
      B_i=\sum_{j=1}^{m_i} b_i^j(w_i^1,\dots,w_i^{m_i},u,\dots,u_{k_i})\frac{\pd}{\pd w_i^j},\quad i=1,2,
\end{gather*}
such that both $A_1$ and $A_2$ satisfy~\er{d=0} 
for all $s\ge 1$ and some point~\er{ua}.
Let 
\begin{equation}
\label{wf}
w^j_2=\varphi^j(w^1_1,\dots,w^{m_1}_1,u,u_1,\dots),
\quad j=1,\dots,m_2, 	
\end{equation}
determine a morphism of these coverings, i.e., 
\begin{gather}
\label{daf}
	(D_x+A_1)\varphi^j=a_2^j(\varphi^1,\dots,\varphi^{m_2},u,\dots,u_{k_2}),\\
\notag	(D_t+B_1)\varphi^j=b_2^j(\varphi^1,\dots,\varphi^{m_2},u,\dots,u_{k_2})
\end{gather}
for all $j=1,\dots,m$. 
Then functions~\er{wf} do not actually depend on any $u_i,\,i\ge 0$.
\end{lemma}
\begin{proof}
Let $r\ge 0$ be the maximal integer such that at least one of functions~\er{wf} 
depends nontrivially on $u_r$. 
Differentiate~\er{daf} with respect to $u_{r+1}$ and substitute $u_i=a_i$ for $i\ge r+1$. 
Taking into account~\er{d=0} for $s=r+1$, we obtain that the right-hand side is zero, 
while on the left-hand side we get $\pd\varphi^j/\pd u_r$. 
Therefore, $\pd\varphi^j/\pd u_r=0$ for all $j$, which 
contradicts to our assumption. 
\end{proof}

\begin{lemma}
\label{lsym}
Consider covering~\er{gc} satisfying~\er{d=0} for all $s\ge 1$ and let 
\begin{equation*}
S=\sum_{j=1}^m s^j(w^1,\dots,w^m,u,u_1,\dots)\frac{\pd}{\pd w^j}	
\end{equation*}
be a symmetry of it.
Then $S$ does not actually depend on any $u_i,\,i\ge 0$.
\end{lemma}
\begin{proof}
Analyzing the equation $[D_x+A,S]=0$ from~\er{symx}, this is proved 
similarly to the previous lemma. 	
\end{proof}

\begin{lemma}
\label{lcir}
Consider covering~\er{gc} satisfying~\er{d=0} for all $s\ge 1$.
Let $\CE'$ be a subequation of $\CE\times W$. Then locally 
$\CE'$ is of the form $\CE_1\times W'$, where $\CE_1$  
is an open subset of $\CE$ and $W'$ is a submanifold of $W$ 
such that all vector fields 
\begin{equation*}
\{A(u,\dots,u_k),\,B(u,\dots,u_k)
\in D(W)\,|\,u,\dots,u_k\in\Com\}	
\end{equation*}
are tangent to $W'$.	
\end{lemma}
\begin{proof}
According to Definition~\ref{subeq}, 
a subequation of codimension $l$
is given by functions 
$$
f_i(w^1,\dots,w^m,u,u_1,\dots),\quad i=1,\dots,l,
$$ 
defined on an open subset $U\subset\CE\times W$
such that 
\begin{itemize}
	\item $f_1(c)=\dots=f_l(c)=0$ for some $c\in U$,
	\item the differentials 
	$$
	d_bf_i\in T^*_b(\CE\times W),\quad i=1,\dots,l,
	$$ 
	are linearly independent for each $b\in U$,
	\item the ideal $I$ of functions on $U$ generated by $f_1,\dots,f_l$ is preserved 
	by the action of the vector fields $D_x+A,\,D_t+B$.
\end{itemize}
Let $z\in W$ be the image of $c$ under the projection $\CE\times W\to W$.
To prove the lemma, it is sufficient to find a 
set of functions 
\begin{equation}
\label{gaa}
g_\alpha(w^1,\dots,w^m),\quad \alpha\in\Lambda,
\end{equation} 
defined on a neighborhood of $z$ such that the ideal of functions 
on a neighborhood $U'\subset U$ of $c$ generated by functions~\er{gaa} 
coincides with $I\bigl|_{U'}$.

Let $r$ be the maximal integer such that at least one of the functions 
$f_1,\dots,f_l$ depends nontrivially on $u_r$. Note that $f_i$ are defined 
on an open subset $V$ of $\Com^{r+1}\times W$ with the coordinates 
$u_0,\dots,u_{r},w^1,\dots,w^m$, the subset 
$$
M=\{q\in V\,|\,f_1(q)=\dots=f_l(q)=0\}
$$ 
is a submanifold of codimension $l$ in $V$, 
and $I\bigl|_V$ coincides with the ideal 
of functions on $V$ that vanish on $M$. 
Thus we essentially have a question of finite-dimensional complex analysis. 

Since 
$$
\frac{\pd}{\pd u_{r+1}}(I)\subset I,\quad (D_x+A)(I)\subset I,
$$
we have
\begin{equation}
\label{arf}
\frac{\pd}{\pd u_{r+1}}\Bigl((D_x+A)(f_i)\Bigl)=\frac{\pd f_i}{\pd u_{r}}
+\frac{\pd A}{\pd u_{r+1}}(f_i)\,\,\in\, I.	
\end{equation}
Substituting $u_i=a_i,\,i\ge r+1$, to~\er{arf}, 
from~\er{d=0} for $s=r+1$ we obtain $\pd f_i/\pd u_{r}\in I$. 
Therefore, the vector field $\pd/\pd u_{r}$ is tangent to $M$, 
which allows   
to generate $I$ on some neighborhood of $c$ 
by functions that do not depend on $u_i$ for $i\ge r$. 
Proceeding by induction on $r$, one completes the proof.
\end{proof}

Applying this lemma to the identity covering $\CE\to\CE$, 
we obtain the following.
\begin{theorem}
Any connected open subset of the translation-invariant 
infinite prolongation of any evolution equation~\er{gev} 
is irreducible. 	
\end{theorem}


Let us introduce some auxiliary notions.
\begin{definition}
\label{perfa}
For each $i\in\zp$, let $V_i$ be a connected open subset of $\Com$ 
such that for all but a finite number of $i$ 
we have $V_i=\Com$. Set 
\begin{equation*}
D=\{(u_0,u_1,\dots,u_i,\dots)\,|\,u_i\in V_i\}.	
\end{equation*}
Let $\CP$ be an algebra of functions on $D$ 
such that each $f\in \CP$ is a complex-analytic 
function dependent on a finite number 
of the variables $u_i,\,i\ge 0$. 
The algebra $\CP$ is said to be \emph{perfect} if for 
each function $f(u_0,\dots,u_r)\in\CP$ and any $i\in\zp$ 
the following conditions hold.
\begin{enumerate}
	\item One has $\pd f/\pd u_i\in \CP$.
	\item
	\label{dg} 
	There is $g(u_0,\dots,u_r)\in \CP$ such that
	$\pd g/\pd u_i=f$.
	\item For any $s<r$ and any fixed numbers $a_i\in V_i,\,i\ge s$, we have 
	$$
	f(u_0,\dots,u_{s-1},a_s,a_{s+1},\dots,a_r)\in \CP.
	$$ 
	\item For all $j\ge 1$ we have $u_j\in \CP$.
\end{enumerate}
Then each function $f\in \CP$ is also called \emph{perfect}.
\end{definition}  

\begin{example}
\label{polpa}
Let $V_i=\Com$ and $\CP$ be the algebra of polynomials 
in $u_i,\,i\ge 0$. Evidently, the algebra $\CP$ is perfect.	
\end{example}

Fix open subsets $V_i\subset\Com$ satisfying the assumptions 
of Definition~\ref{perfa} and a perfect algebra $\CP$.

\begin{definition}
Consider a vector field 
\begin{equation}
\label{awu}
A=\sum_{j=1}^m a^j(w^1,\dots,w^m,u,u_1,\dots,u_k)\frac{\pd}{\pd w^j}	
\end{equation}
defined on an open subset of $W\times V_0\times\dots\times V_k$.

A vector field 
\begin{equation}
\label{sw}
S=\sum_{j=1}^m s^j(w^1,\dots,w^m)\frac{\pd}{\pd w^j}	
\end{equation}
is said to be \emph{$1$-primitive \textup{(}with respect to $A$\textup{)}} 
if $[S,A]=0$. 
Now by induction on $q\in\mathbb{N}$  
a vector field~\er{sw} is called 
\emph{$q$-primitive \textup{(}with respect to $A$ and $\CP$\textup{)}}
if the commutator $[S,A]$ can be presented as a sum $\sum_{j=1}^Nf_jS_j$,
where $S_j$ are $(q-1)$-primitive fields and $f_j$ 
are perfect functions. 
In particular, one has $(\mathrm{ad}^qA)(S)=0$. 

A vector field  
\begin{equation}
\label{swu}
S=\sum_{j=1}^m s^j(w^1,\dots,w^m,u,u_1,\dots)\frac{\pd}{\pd w^j}	
\end{equation}
is said to be \emph{primitive}
(without any prefix) if one has $S=\sum_{j=1}^Nf_jS_j$,
where $f_j$ are perfect functions and $S_j$ are $q$-primitive 
vector fields for some $q$.  
\end{definition}

\begin{remark}
Below all primitive vector fields are 
primitive with respect to $A$ and $\CP$, 
where $\CP$ is a fixed perfect algebra and 
$A$ arises from a covering~\er{gc}.

Evidently, 
primitive vector fields form a module over the algebra $\CP$.	
\end{remark}
  
\begin{lemma}
\label{nd} 
Consider an arbitrary vector field~\er{swu} 
defined on a neighborhood of the point $u_i=a_i\in V_i,\,i\ge 0$.  
Consider a covering~\eqref{gc} satisfying~\eqref{d=0} for all $s\ge 1$.
\begin{enumerate}
	\item \label{xxa}
	If 
\begin{equation} 
\label{xxae}
	\frac{\pd}{\pd u_i}\Bigl(D_x(S)+[A,S]\Bigl)=0\quad \forall\,i>0
\end{equation}
then $\pd S/\pd u_i=0$ for all $i\ge 0$. 
 \item
 \label{pxa}
 If $D_x(S)+[A,S]$ is primitive then $S$ is primitive.
 \item
 \label{pxb}
 If $S$ is primitive and the function $F$ in~\er{gev} is perfect 
 then $[B,S]$ is primitive.
\end{enumerate}
\end{lemma}
\begin{proof}
(1) Let $r$ be the maximal integer such that $\pd S/\pd u_r\neq 0$.
From~\er{d=0} for $s=r+1$ we have 
\begin{equation}
\label{sur}
	\frac{\pd}{\pd u_{r+1}}\Bigl(D_x(S)+[A,S]\Bigl)(u,\dots,u_r,a_{r+1},\dots,a_k)=
	\frac{\pd S}{\pd u_{r}}.
\end{equation}
Combining this with~\er{xxae} for $i=r+1$, we obtain $\pd S/\pd u_r=0$.

(2) 
Again let $r$ be the maximal integer such that $\pd S/\pd u_r\neq 0$. 
Then~\er{sur} holds. Since $D_x(S)+[A,S]$ is primitive, 
vector field~\er{sur} is also primitive, by the properties of perfect functions.
Therefore, by Condition~\ref{dg} of Definition~\ref{perfa}, 
there is a primitive field $S'$ such that $\tilde S=S-S'$ does not depend 
on $u_i,\,i\ge r$. Then $D_x(\tilde S)+[A,\tilde S]$ is primitive, and 
by induction on $r$ one completes the proof.


(3) Applying $\mathrm{ad}\,S$ to~\er{gc}, we obtain
\begin{equation}
\label{gcads}
D_x\bigl([S,B]\bigl)+[A,[S,B]]=[S,D_tA]-[[S,A],B].	
\end{equation}
By assumption, for some $q$ one has
\begin{equation}
\label{sfj}
	S=\sum_jf_jS_j,\quad f_j\text{ are perfect},\quad 
	S_j\text{ are $q$-primitive}.  
\end{equation}
Let us prove that $[S,B]$ is primitive by induction on $q$.
For $q=1$ the right-hand side of~\er{gcads} is zero. 
Applying Part~\ref{xxa} of this lemma to the vector field $[S,B]$, 
we obtain that $[S,B]$ is 1-primitive.
 
Now assume that the statement holds for $q-1$.
Consider an arbitrary vector field $S$ satisfying~\er{sfj}.
Let us prove that $[S,B]$ is primitive.

By formula~\er{dt}, we have 
\begin{equation}
\label{sda}
	[S,D_tA]=\sum_{j=0}^kD_x^j(F)[S,\frac{\pd}{\pd u_j}A].
\end{equation}
Since $F$ is perfect, the functions $D_x^j(F)$ 
are also perfect. Besides, for any primitive $X$ 
the vector fields $[X,{\pd A}/{\pd u_j}]$ are also primitive 
for all $j$. Therefore,~\er{sda} is primitive.
 
Since $[S,A]$ is a linear combination of $(q-1)$-primitive fields, 
the vector field $[[S,A],B]$ is also primitive 
by the induction assumption.
Thus the right-hand side of~\er{gcads} is primitive 
and we can apply Part~\ref{pxa} of this lemma to $[S,B]$. 
\end{proof}

\section{Coverings of the KdV equation}

\label{sec_ckdv}


In this section we return to the KdV equation
\begin{equation}
\label{kdv}
        u_t=u_3+u_1u.
\end{equation}
Our final goal here is Theorem~\ref{mtkdv}. 

\subsection{The canonical form of coverings}

\label{cfkdv}


\begin{theorem}
\label{cp}
For any covering of equation~\eqref{kdv}
\begin{gather}
\label{ck}
        D_xB-D_tA+[A,B]=0,\\
\label{depend}\notag
        A=A(u,u_1,\dots,u_{k}),\quad
        B=B(u,u_1,\dots,u_k)
\end{gather}
\textup{(}we omit the dependence on fibre coordinates $w^j$\textup{)}
locally there is an equivalent covering such that $A,\,B$ 
are polynomial in $u_i$ and $A$ satisfies~\er{d=0} for all $s\ge 1$ 
and $a_i=0,\,i\ge 1$. 
\end{theorem}

\begin{proof}
Consider an arbitrary point 
$u_i=a_i\in\Com,\,w^j=w^j_0\in\Com$ where the vector fields $A$ and $B$ 
are defined. All local gauge transformations 
in this proof will be defined on a neighborhood of this point.  
By Lemma~\ref{gcf}, we can assume that~\eqref{d=0} holds 
for all $s\ge 1$. 
\begin{remark}
\label{extend}
It would be most convenient 
to take $a_i=0$ from the beginning.
However, since we consider coverings over arbitrary open subsets 
of the translation-invariant infinite prolongation of~\er{kdv}, 
we do not know in advance whether $A,\,B$ are defined around this point. 
We will show by induction that after a suitable gauge transformation 
the vector fields $A,\,B$ become polynomial in $u_i$ and, therefore, 
are uniquely extended to the whole space of variables $u,\dots,u_k$.
\end{remark}


To clarify further arguments, 
let us first determine the form of $A,\,B$ with respect to 
the highest derivatives $u_i,\,i\ge k-3$. 
A straightforward 
analysis of equation~\eqref{ck} 
shows that $A$ does not depend on $u_k,\,u_{k-1}$ and is a polynomial of degree 
$2$ in $u_{k-2}$, while $B$ is polynomial in $u_k,\,u_{k-1},\,u_{k-2}$. 
Therefore, following the strategy of Remark~\ref{extend}, we can find a gauge transformation 
such that the transformed $A$ satisfies~\eqref{d=0} with $a_i=0$ for $i\ge k-2$. 

Then~\eqref{d=0} for $s=k-2$ implies
\begin{equation}
\label{A}
        A=u_{k-2}^2A_2(u,\dots,u_{k-3})+A_0(u,\dots,u_{k-3}).
\end{equation} 
Further analysis shows that $A_2$ does not depend on $u_{k-3}$ 
and $B$ is of the form 
\begin{multline}
\label{B}
B=2u_{k-2}u_kA_2-u_{k-1}^2A_2+B_{11}(u,\dots,u_{k-3})u_{k-2}u_{k-1}+\\
+B_{10}(u,\dots,u_{k-3})u_{k-1}+B_0(u,\dots,u_{k-2}).
\end{multline}

Differentiating~\eqref{ck} with respect to $u_k,\,u_{k-2}$, we obtain 
\begin{equation*}
\label{m1}
    2D_x(A_2)+B_{11}+2[A_0,A_2]=0,
\end{equation*}
while differentiation with respect to $u_{k-1},\,u_{k-1}$ implies
\begin{equation*}
  -D_x(A_2)+B_{11}-[A_0,A_2]=0.
\end{equation*}
Therefore,  
\begin{equation}
\label{dxaa2}
	D_x(A_2)+[A,A_2]=0,
\end{equation}
which by Lemma~\ref{nd} (\ref{xxa}) says that $A_2$ does not 
depend on $u_i,\,i\ge 0$, and $[A,A_2]=0$. 
That is, $A_2$ is $1$-primitive with respect to $A$.


\begin{definition}
Let $r\in\zp$ and $r<k$.
A vector field
\begin{equation*}
A=\sum_{j=1}^m a^j(w^1,\dots,w^m,u,u_1,\dots,u_k)\frac{\pd}{\pd w^j}	
\end{equation*}
is said to be \emph{$r$-simple} 
if it satisfies~\er{d=0} for all $s\ge 1$ with  
$a_i=0,\,i\ge k-r$, and some $a_1,\dots,a_{k-r-1}\in\Com$. 
\end{definition}

\begin{lemma}
\label{simprim}
\begin{enumerate}
\item 
For each $r<k$ and any covering~\er{ck} 
there is a locally gauge equivalent 
covering with $r$-simple $A$.
\item
If a covering~\er{ck} has $r$-simple $A$ 
then the vector fields
\begin{equation}
\label{aabb}
\begin{aligned}
A'&=A(u,\dots,u_k)-A(u,\dots,u_{k-r-1},0,\dots,0),\\
B'&=B(u,\dots,u_k)-B(u,\dots,u_{k-r+1},0,\dots,0)
\end{aligned}
\end{equation}
are primitive with respect to $A$ and $\CP$, 
where $\CP$ is the perfect algebra 
constructed in Example~\ref{polpa}.	
\end{enumerate}
\end{lemma}
\begin{proof}
For $r=2$ we proved these statements above. 
Suppose that the statements of hold for some $r=l\le k-2$ 
and let us prove them for $r=l+1$.

By assumption, each covering is locally equivalent 
to a covering~\er{ck} with $l$-simple $A$. 
Then by Part~2 of the lemma we have 
\begin{equation}
\label{abr}
\begin{aligned}
A&=A'(u,\dots,u_{k-2})+\tilde A(u,\dots,u_{k-l-1}),\\
B&=B'(u,\dots,u_{k})+\tilde B(u,\dots,u_{k-l+1}),	
\end{aligned}
\end{equation}
where 
\begin{equation}
\tilde A=A(u,\dots,u_{k-l-1},0,\dots,0),\quad 
\tilde B=B(u,\dots,u_{k-l+1},0,\dots,0),	
\end{equation}
and the primitive vector fields $A',\,B'$ 
are given by~\er{aabb} for $r=l$. 

We can rewrite~\er{ck} as follows 
\begin{equation}
\label{ckp}
D_x \tilde B-D_t\tilde A+[\tilde A,\tilde B]+P=0,	
\end{equation}
where 
\begin{equation}
\label{pba}
P=D_xB'-D_tA'+[A,B']+[A',B]	
\end{equation}
is primitive. Indeed, the fact that 
$D_xB',\,D_tA',\,[A,B']$ are primitive follows 
immediately from the fact that $A',\,B'$ are primitive,
while $[A',B]$ is primitive by Lemma~\ref{nd}~\er{pxb}. 
In particular, $P$ is polynomial in $u_i,\,i\ge 0$.

From equation~\er{ckp} it follows easily that 
$\tilde A$, $\tilde B$ are polynomial in 
$u_{k-l-1}$, $u_{k-l}$, $u_{k-l+1}$.
Therefore, $A(u,\dots,u_k)$ and $B(u,\dots,u_k)$ 
are defined for $u_i=a_i,\,i\le k-l-2$, and arbitrary values 
of $u_j,\,j\ge k-l-1$. 
By Lemma~\ref{gcf}, after some gauge transformation
\begin{equation*}
w^i\mapsto g^i(w^1,\dots,w^m,u,\dots,u_{k-l-2})	
\end{equation*}
$A$ becomes $(l+1)$-simple, 
which proves Part~1 of the lemma for $r=l+1$.

To prove Part~2, consider an arbitrary 
covering~\er{ck} with $(l+1)$-simple $A$, where $l\le k-2$.
Since $(l+1)$-simple $A$ is also $l$-simple, 
we again have representation~\er{abr} and 
equation~\er{ckp}, where~\er{pba} is primitive.

Similarly to formulas~\er{A} and~\er{B}, from~\er{ckp} 
we obtain 
\begin{gather}
\label{Ag}
	 \tilde A=P_1+u_{k-l-1}^2A'_2(u,\dots,u_{k-l-3})+A'_0(u,\dots,u_{k-l-2}),\\
\label{Bg}	 
     \tilde B=P_2+2u_{k-l-1}u_{k-l+1}A'_2-u_{k-l}^2A'_2+
      B'_{11}(u,\dots,u_{k-l-2})u_{k-l-1}u_{k-l}+\\
      \notag
        +B'_{10}(u,\dots,u_{k-l-2})u_{k-l}+B'_0(u,\dots,u_{k-l-1}),
\end{gather}
where $P_1,\,P_2$ are primitive.
Similarly to~\er{dxaa2}, this implies that 
$D_xA'_2+[A,A'_2]$ is also primitive. 
By Lemma~\ref{nd}~\er{pxa}, 
the vector field $A'_2$ is primitive.

Then the vector fields
\begin{gather*}
A(u,\dots,u_k)-A(u,\dots,u_{k-l-2},0,\dots,0)
,\\
B(u,\dots,u_k)-B(u,\dots,u_{k-l},0,\dots,0)
\end{gather*}
are also primitive, which proves Part~2 of the lemma for $r=l+1$.
\end{proof}

By the above lemma for $r=k-1$, we obtain that 
after a suitable gauge transformation one has 
\begin{gather*}
A=A''(u,\dots,u_{k-2})+A''_0(u),\\
B=B''(u,\dots,u_{k})+B''_0(u,u_1,u_2),
\end{gather*}
where $A'',\,B''$ are primitive and $A$ is $(k-1)$-simple. 
Now it is straightforward to prove that 
$A_0'',\,B_0''$ are polynomial in $u,\,u_1,\,u_2$.
\end{proof}


\subsection{The fundamental algebras}
\label{cfakdv}

From the above proof it follows that for each $k\ge 3$
there are finite subsets 
$$
\CM_k\subset\zp^{k-1},\quad\CN_k\in\zp^{k+1}
$$ 
such that the following statement holds.
If a covering~\er{ck} of equation~\er{kdv} 
satisfies~\er{d=0} for all $s\ge 1$ with $a_i=0,\,i\ge 1$, 
then it is of the form
\begin{equation}
\label{abp}
\begin{aligned}
A&=\sum_{(i_0,\dots,i_{k-2})\in \CM_k}u_0^{i_0}\dots u_{k-2}^{i_{k-2}}\ca_{i_0\dots i_{k-2}},\\
B&=\sum_{(i_0,\dots,i_k)\in \CN_k}u_0^{i_0}\dots u_k^{i_k}\cb_{i_0\dots i_k},		
\end{aligned}
\end{equation}
where the vector fields
\begin{equation}
\label{abi}
\ca_{i_0\dots i_{k-2}},\quad \cb_{i_0\dots i_k}	
\end{equation}
do not depend on $u_i,\,i\ge 0$.

Let us show that this canonical form of coverings satisfies 
Definition~\ref{pfa} if we take 
\begin{gather*}
\CA_k=\{u_0^{i_0}\dots u_{k-2}^{i_{k-2}}\,|\,(i_0,\dots,i_{k-2})\in \CM_k\},\\	
\CB_k=\{u_0^{i_0}\dots u_{k}^{i_{k}}\,|\,(i_0,\dots,i_{k})\in \CN_k\}.
\end{gather*}

Relation~\er{abin} is obvious. 
Condition~\ref{cmain} of 
Definition~\ref{pfa} follows from Theorem~\ref{cp}. Let 
\begin{equation*}
S=\sum_{j=1}^m s^j(w^1,\dots,w^m,u,u_1,\dots)\frac{\pd}{\pd w^j}	
\end{equation*}
be a symmetry of the covering given by vector fields~\er{abp}, i.e.,
\begin{equation}
\label{xts}
	[D_x+A,S]=[D_t+B,S]=0.
\end{equation}
By Lemma~\ref{lsym}, $S$ does not depend on $u_i,\,i\ge 0$. 
Then~\er{xts} implies that $S$ commutes with 
all vector fields~\er{abi}, which proves Condition~\ref{csym} of Definition~\ref{pfa}.
Conditions~\ref{cm} and~\ref{cir} follow analogously
from Lemmas~\ref{lcm} and~\ref{lcir} respectively.

Vector fields~\er{abp} satisfy~\er{ck} if and only if certain 
Lie algebra relations hold for~\er{abi}. 
Denote by $\mf_{k-2}$ the quotient of the free Lie algebra 
generated by letters~\er{abi} over these relations. 
We obtain the system of fundamental algebras
\begin{equation}
\label{fkkdv}
\dots\to\mf_{k+1}\to\mf_k\to\dots\to\mf_1\to\mf_0	
\end{equation}
for equation~\er{kdv}. 
In particular, the algebras $\mf_1$ and $\mf_0$ are described in Example~\ref{fkdv1}.

Denote by $\ma_{k-2}$ the subalgebra of $\mf_{k-2}$ generated 
by $\ca_{i_0,\dots,i_{k-2}}$. 

\begin{lemma}
\label{lbia}
We have
\begin{equation}
\label{bia}
\cb_{i_0\dots i_k}\in\ma_{k-2}\quad\text{for }\,i_0+\dots+i_k>0.	
\end{equation}	
\end{lemma}
\begin{proof}
For $(i_0,\dots, i_k)\in \CN_k$ set 
$$
r(i_0,\dots, i_k)=\max\{s\,|\,i_s>0\}. 
$$	
Let us prove~\er{bia} by induction on $k-r(i_0,\dots, i_k)$. 

For $(i_0,\dots, i_k)\in \CN_k$ with $r(i_0,\dots, i_k)=k$
it follows from~\er{B}.   

Suppose that~\er{bia} holds for all $(i_0,\dots, i_k)\in \CN_k$
with $r(i_0,\dots, i_k)\ge l+1$. Differentiate~\er{ck} with respect to $u_{l+1}$
and substitute $u_i=0$ for $i\ge l+1$. Since $A$ satisfies~\er{d=0} 
for $s=l+1$ and $a_i=0$, we obtain~\er{bia} for 
$(i_0,\dots, i_k)\in \CN_k$ with $r(i_0,\dots, i_k)=l$.
\end{proof}
Combining~\er{bia} and~\er{ck}, one gets 
\begin{equation}
\label{b0a}
[\cb_{0\dots 0},\ma_{k-2}]\subset\ma_{k-2}.
\end{equation}
Substituting $u_i=0$ to~\er{ck}, we obtain also
\begin{equation}
\label{ab0}
[\ca_{0\dots 0},\cb_{0\dots 0}]=0.	
\end{equation}

Let us specify the structure of~\er{abp}. 
For $k=3$ it was described in Theorem~\ref{c1}.
Similarly to the proof of Theorem~\ref{cp}, 
one obtains that for $k\ge 4$ vector fields~\er{abp} have the form
\begin{gather}
\label{asp}
A=\ca_{k-2}\bigl(u_{k-2}^2-\frac{2k-3}3uu_{k-2}^2\bigl)+\ca'_{k-2}u_{k-3}^2+A_0(u,\dots,u_{k-4}),\\	
\label{bsp}
B=2u_{k-2}u_k\ca_{k-2}+B_0(u,\dots,u_{k-1}),
\end{gather}
where 
\begin{gather}
\notag
\ca_{k-2}=\ca_{0\dots 02},\quad \ca'_{k-2}=\ca_{0\dots 020},\\
\label{aka}
[\ca_{k-2},A]=0,\\
\label{akb}
[\ca_{k-2},B]=3[A_0,\ca'_{k-2}].	
\end{gather}
and $A_0,\,B_0$ are polynomial in $u_i$. 

Equation~\er{aka} implies 
\begin{equation}
	\label{aac}
	[\ca_{k-2},\ma_{k-2}]=0.
\end{equation}
Combining this with~\er{bia} and~\er{akb}, we obtain
\begin{equation}
\label{akb0}
	[\ca_{k-2},\cb_{0\dots 0}]=3[\ca_{0\dots 0},\ca'_{k-2}].
\end{equation}
Moreover, taking into account~\er{ab0} and applying 
$\mathrm{ad}^{s}\cb_{0\dots 0}$ to~\er{akb0}, we obtain
\begin{equation}
\label{akbs}
	-(\mathrm{ad}^{s+1}\cb_{0\dots 0})(\ca_{k-2})=
	3[\ca_{0\dots 0},(\mathrm{ad}^{s}\cb_{0\dots 0})(\ca'_{k-2})]\quad
	\forall\,s\ge 0.
\end{equation}

By the definition of $\mf_n$ 
and formulas~\er{asp},~\er{bsp}, for each $n\ge 2$
the algebra $\mf_{n-1}$ is isomorphic to the quotient 
of $\mf_n$ over the ideal $\mi_n$ generated by $\ca_{n}$. 
From~\er{bia},~\er{b0a}, and~\er{akbs} we obtain 
that $\mi_n\subset\ma_{n}$. 
Moreover,~\er{aac} implies 
\begin{equation}
\label{ia0}
[\mi_n,\ma_{n}]=0.	
\end{equation}

\begin{lemma}
For each $n\ge 1$ we have the relation
\begin{equation}
\label{baak0}
	-(\mathrm{ad}^{n}\cb_{0\dots 0})(\ca_{n})=0
\end{equation}	
in the algebra $\mf_{n}$.
\end{lemma}
\begin{proof}
For $n=1$ this statement follows from~\er{crel1}. 
By induction on $n$, 
suppose that~\er{baak0} holds for $n-1$. 
By formula~\er{asp}, the generator $\ca'_n\in\mf_n$ is mapped to 
$\ca_{n-1}\in\mf_{n-1}$ by the natural epimorphism
$\mf_n\to\mf_{n}/\mi_n\cong\mf_{n-1}$. 
Therefore, 
$(\mathrm{ad}^{n-1}\cb_{0\dots 0})(\ca'_n)\in\mi_n$.	
Combining this with~\er{akbs} and~\er{ia0}, we obtain 
\er{baak0}.
\end{proof}
From the above results it follows that the elements
$$
c_i=(\ad^i\cb_{0\dots 0})(\ca_{n}),\quad 
i=0,\dots,n-1,
$$
span the ideal $\mi_n$.
The element $c_{n-1}$ belongs to the center of $\mf_n$. 
Moreover, for each $i=0,\dots,n-1$ 
the image of $c_{i}$ belongs to the center of the quotient
\begin{equation*}
	\mf_n/\langle c_{i+1},\dots,c_{n-1}\rangle.
\end{equation*}
Thus we have the following statement.
\begin{lemma}
\label{ff}
For each $n\ge 2$ the algebra $\mf_n$ is obtained from 
$\mf_{n-1}$ applying the operation of one-dimensional 
central extension no more than $n$ times. 	
\end{lemma}

Let us now prove the main result of this section.

\begin{theorem}
\label{mtkdv}
The KdV equation~\er{kdv} possesses fundamental 
algebras~\er{fkkdv}. Each algebra $\mf_k$ is quasi-finite and is obtained from 
the algebra $\sl_2(\mathbb{C})\otimes_{\mathbb{C}}\Com[\lambda]$ 
applying several times the operation 
of one-dimensional central extension. 
\end{theorem}
\begin{proof}
It was shown above that~\er{fkkdv} are fundamental algebras of~\er{kdv}.
Let us prove that algebras~\er{fkkdv} are quasi-finite.

By Theorem~\ref{slih}, the algebra 
$$
\mg=\sl_2(\mathbb{C})\otimes_{\mathbb{C}}\Com[\lambda]
$$ 
is quasi-finite. 
From~\er{crel1} it follows that $\mf_1$ is the trivial 
central extension of the algebra $\mL$ from Proposition~\ref{eck}.

Since the Heisenberg algebra $H$ is nilpotent, 
the algebra $\mf_1$ is obtained 
from $\mg$ applying 6 times the operation of one-dimensional 
central extension. Therefore, by Theorem~\ref{qfce}, the algebra
$\mf_1$ is also quasi-finite. Finally, combining Lemma~\ref{ff} and Theorem~\ref{qfce},
we obtain that all fundamental algebras~\er{fkkdv} are quasi-finite. 
\end{proof}

It is well known that $\sl_2(\mathbb{C})\otimes_{\mathbb{C}}\Com[\lambda]$ 
has no nontrivial central extensions. Combining this with Theorem~\ref{mtkdv}, 
we obtain the following specification of the structure of $\mf_k$.

\begin{theorem}
Each algebra $\mf_k$ is isomorphic to the direct sum of 
$\sl_2(\mathbb{C})\otimes\Com[\lambda]$ 
and a finite-dimensional nilpotent algebra.	
\end{theorem}

\section{Coverings of the Krichever-Novikov equation} 
\label{sec_ckn}

Consider the Krichever-Novikov (KN) equation \cite{krich,kncl,sok2}
\begin{equation}
  \label{kn}
 u_t=u_3-\frac32\frac{u_2^2}{u_1}+
 \frac{h(u)}{u_1},\quad
 u_k=\frac{\partial^k u}{\partial x^k}, 
\end{equation}
where $h(u)$ is a polynomial of degree $3$ with coefficients in $\Com$. 
If the roots of the polynomial $h(u)$ are
distinct then equation~\eqref{kn} is said to be \emph{nonsingular}.

The main goal of this section is Theorem~\ref{fakn}.

\subsection{The canonical form of coverings}

We want to have an analog of Theorem~\ref{cp} for equation~\er{kn}.
The straightforward repetition of the proof of Theorem~\ref{cp} 
is not possible, because~\er{kn} is not polynomial in $u_1$. 

To overcome this, we need to introduce a perfect algebra  
that contains the function $1/u_1$. 
By Condition~\ref{dg} of Definition~\ref{perfa}, 
this algebra must contain also $\int 1/u_1du_1$. 

To this end, 
choose a half-line $L\subset\Com$ from $0$ to $\infty$ 
such that $V_1=\Com\setminus L$ is simply connected. 
Let $\ln u_1$ be a single-valued branch of the logarithm 
defined on $V_1$. Set $V_i=\Com,\,i\neq 1$, and let 
$\CP$ be the algebra of polynomials in
\begin{equation}
\label{polu}
u_i,\ i\ge 0,\quad \frac{1}{u_1},\quad
\ln u_1.	
\end{equation}
Then $\CP$ is a perfect algebra. 
Indeed, all conditions of Definition~\ref{perfa} 
are obvious except of Condition~\ref{dg}. 
The latter follows from the fact that 
for any $a\in\mathbb{Z},\,b\in\zp$ 
there is a polynomial $g$ in $u_1,\,1/u_1,\,\ln u_1$ 
such that $\pd g/\pd u_1=u_1^a\ln^b u_1$.	

\begin{remark}
Thus for equation~\er{kn} we study 
not the whole translation-invariant infinite prolongation, but the open dense subset 
\begin{equation*}
\{(u_0,u_1,\dots)\,|\ u_1\in\Com\setminus L,\quad u_i\in\Com\quad \forall\,i\neq 1\}	
\end{equation*}
of it.
\end{remark}

In Theorem~\ref{cp}, we proved that every covering 
of the KdV equation is locally equivalent to a covering 
in the canonical form satisfying~\er{d=0} for all $s\ge 1$ 
and $a_i=0,\,i\ge 1$. 
For equation~\er{kn} the point $u_i=0$ 
is also crucial.
However, one cannot prove the same statement 
for coverings of~\er{kn},  
because $1/u_1$ and $\ln u_1$ are not defined at $u_1=0$.
Let us make necessary modifications.

\begin{definition}
A vector field 
\begin{equation}
\label{swukn}
S=\sum_{j=1}^m s^j(w^1,\dots,w^m,u,\dots,u_k)\frac{\pd}{\pd w^j}	
\end{equation}
is said to be \emph{$u_1$-free} if each function 
$s^j(w^1,\dots,w^m,u,\dots,u_k)$ is polynomial in
\begin{equation}
\label{var1}
u_i,\ i\ge 1,\quad \frac{1}{u_1},\quad
\ln u_1	
\end{equation}
with coefficients dependent on $u,\,w^1,\dots,w^m$ 
and the coefficient at the monomial $u_1$ is zero.
(This coefficient is well defined because the functions 
$$
u_1^a\ln^b u_1,\quad a\in\mathbb{Z},\quad b\in\zp, 
$$
are linearly independent.) 
\end{definition}

\begin{definition}
\label{wsimple}
Let $r\in\zp$ and $r\le k-2$.
A vector field
\begin{equation*}
A=\sum_{j=1}^m a^j(w^1,\dots,w^m,u,u_1,\dots,u_k)\frac{\pd}{\pd w^j}	
\end{equation*}
is said to be \emph{weakly $r$-simple} 
if it satisfies~\er{d=0} for all $s\ge 2$ with  
$a_i=0,\,i\ge k-r$, and some $a_2,\dots,a_{k-r-1}\in\Com$. 
\end{definition}

In contrast to $r$-simple vector fields, a weakly $r$-simple 
vector field does not necessarily satisfy~\er{d=0} 
for $s=1$.

\begin{remark}
In this section perfect functions are elements of 
the perfect algebra $\CP$ defined above.	
\end{remark}
  
\begin{lemma}
\label{repl}
If in Lemmas~\ref{lcm},~\ref{lsym},~\ref{lcir},~\ref{nd} one replaces the 
condition that $A$ satisfies~\er{d=0} for all $s\ge 1$
by the condition that $A$ is $u_1$-free and weakly $(k-2)$-simple 
then the conclusions of these lemmas remain valid.
\end{lemma}
\begin{proof}
Let us prove that Lemma~\ref{nd}~\er{xxa} remains valid, 
since the other statements are proved analogously.

So assume that $A(u,\dots,u_k)$ is $u_1$-free and weakly $(k-2)$-simple 
and that equation~\er{xxae} holds. 
By Definition~\ref{wsimple},  
$A$ satisfies~\er{d=0} for all $s\ge 2$. 
Therefore, the equations 
\begin{equation}
\label{su2}
\frac{\pd S}{\pd u_i}=0\quad
\forall\,i\ge 1	
\end{equation}
are proved in the same way as in Lemma~\ref{nd}~\er{xxa}.

Let us prove that $\pd S/\pd u$ is also equal to zero. 
From~\er{xxae} for $i=1$ we have 
\begin{equation}
\label{su1a}
\frac{\pd S(u)}{\pd u}+[\frac{\pd A}{\pd u_1},S(u)]=0.	
\end{equation}
Since $A$ is $u_1$-free and~\er{su2} holds, 
the vector field $[{\pd A}/{\pd u_1},S(u)]$ 
is either zero or depends nontrivially on some $u_i,\,i\ge 1$.
Combining this with~\er{su2} and~\er{su1a}, we obtain 
$$
\frac{\pd S}{\pd u}=[\frac{\pd A}{\pd u_1},S(u)]=0.
$$
\end{proof}


\begin{theorem}
\label{cpkn}
For any covering of equation~\eqref{kn}
\begin{gather}
\label{ckkn}
        D_xB-D_tA+[A,B]=0,\\
\label{dependkng}
        A=A(u,u_1,\dots,u_{k}),\quad
        B=B(u,u_1,\dots,u_k)
\end{gather}
\textup{(}we omit the dependence on fibre coordinates $w^j$\textup{)}
locally there is an equivalent covering such that 
\begin{enumerate}
	\item $A,\,B$ are polynomial in~\er{polu},
	\item $A$ is $(k-2)$-simple and $u_1$-free.
\end{enumerate}
\end{theorem}
\begin{proof}
Let~\er{dependkng} be defined on a neighborhood of a point 
$u_i=a_i$. 


\begin{lemma}
\label{simprimkn}
\begin{enumerate}
\item 
For each $r\le k-2$ and any covering~\er{ckkn} 
there is a locally gauge equivalent 
covering with $r$-simple $A$.
\item \label{s-p}
If a covering~\er{ckkn} has $r$-simple $A$ 
then the vector fields
\begin{equation*}
\begin{aligned}
A'&=A(u,\dots,u_k)-A(u,\dots,u_{k-r-1},0,\dots,0),\\
B'&=B(u,\dots,u_k)-B(u,\dots,u_{k-r+1},0,\dots,0)
\end{aligned}
\end{equation*}
are primitive with respect to $A$ and $\CP$. 
\end{enumerate}
\end{lemma}
\begin{proof}
This is proved similarly to Lemma~\ref{simprim}.
Formulas~\er{Ag} and~\er{Bg} for $l\le k-3$ are replaced by 
\begin{gather*}
	 \tilde A=P_1+\frac{u_{k-l-1}^2}{u_1^2}A'_2+A'_0(u,\dots,u_{k-l-2}),\\
     \tilde B=P_2+2\frac{u_{k-l-1}u_{k-l+1}}{u_1^2}A'_2-\frac{u_{k-l}^2}{u_1^2}A'_2+
      B'_{11}(u,\dots,u_{k-l-2})u_{k-l-1}u_{k-l}+\\
        +B'_{10}(u,\dots,u_{k-l-2})u_{k-l}+B'_0(u,\dots,u_{k-l-1}),
\end{gather*}
where $P_1,\,P_2,\,A'_2$ are primitive.
\end{proof}

By the above lemma for $r=k-2$, 
after a suitable gauge transformation we have 
\begin{equation}
\label{abkn}
\begin{aligned}
A&=A'+A_0(u,u_1),\\
B&=B'+B_0(u,u_1,u_2,u_3),	
\end{aligned}
\end{equation}
where the vector fields
\begin{equation}
\label{aabbkn}
\begin{aligned}
A'&=A(u,\dots,u_k)-A(u,u_1,0,\dots,0),\\
B'&=B(u,\dots,u_k)-B(u,u_2,u_3,0,\dots,0)
\end{aligned}
\end{equation}
are primitive and $A$ is $(k-2)$-simple. 

Substituting~\er{abkn} to~\er{ckkn}, 
it is straightforward to obtain that 
\begin{equation}
\label{a0c}
A_0=C+\frac{1}{u_1}A_1(u)+u_1A_2(u)+A_3(u),	
\end{equation}
where $C$ is primitive. 

The vector field $A$ remains 
weakly $(k-2)$-simple 
and polynomial in~\er{var1} after any   
gauge transformation of the form
\begin{equation}
\label{gtu1}
	w^i\mapsto f^i(w^1,\dots,w^m,u).
\end{equation}

Let us find a gauge transformation~\er{gtu1} 
such that $A$ becomes $u_1$-free. 
To this end, let 
$$
\sum_{j=1}^m c^j(w^1,\dots,w^m,u)\frac{\pd}{\pd w^j}
$$
be the coefficient of $A$ at the monomial $u_1$
and consider the system of ordinary 
differential equations
\begin{equation*}
\frac{d}{du}f^j(w^1,\dots,w^m,u)=
c^j(f^1,\dots,f^m,u),\quad j=1,\dots,m,	
\end{equation*}
dependent on the parameters $w^1,\dots,w^m$.
Its local solution with the initial condition  
$$
f^j(w^1,\dots,w^m,a_0)=w^j,\quad j=1,\dots,m, 
$$
determines gauge transformation~\er{gtu1} 
that makes $A$ $u_1$-free.  

By Lemma~\ref{repl}, in Lemma~\ref{simprimkn}\er{s-p} 
for $r=k-2$ the condition that $A$ is $(k-2)$-simple 
can be replaced by the condition that 
$A$ is weakly $(k-2)$-simple and $u_1$-free.
Therefore, after this gauge transformation 
vector fields~\er{aabbkn} remain primitive 
and we have formula~\er{a0c} with primitive $u_1$-free $C$ 
and $A_2(u)=0$. 

Now it is straightforward to show that 
\begin{equation}
A(u,u_1,0,\dots,0),\quad
B(u,u_2,u_3,0,\dots,0)
\end{equation}
are polynomial in~\er{polu}. 
Therefore, $A$ and $B$ satisfy the conditions 
of the theorem.
\end{proof}

\subsection{The fundamental algebras}
\label{cfakn}

Consider the following set of perfect functions 
\begin{multline*}
Z=\{(\ln^a u_1)u_0^{i_0}u_1^{i_1}u_2^{i_2}\dots u_k^{i_k}\,|\,
i_1\in\mathbb{Z},\\
a,i_0,i_2,\dots,i_k\in\zp,\,|i_1|+a+i_0+i_2+\dots+i_k> 0\}.	
\end{multline*}
Similarly to the case of the KdV equation,
from the proof of Theorem~\ref{cpkn} 
it follows that for each $k\ge 3$
there are finite subsets  
$$
\CA'_k,\,\CB'_k\subset Z,\quad 
\CA'_k\subset\CA'_{k+1},\quad 
\CB'_k\subset\CB'_{k+1}
$$ 
such that the following statement holds.
If a covering~\er{ckkn} of equation~\er{kn} 
has $(k-2)$-simple $u_1$-free $A$ then it is of the form
\begin{equation}
\label{abpkn}
A=\sum_{f\in\CA'_k}f\ca_f+\ca_1,\quad 
B=\sum_{g\in\CB'_k}g\cb_g+\cb_1,	
\end{equation}
where the vector fields
\begin{equation}
\label{abikn}
\ca_f,\ \cb_g,\ \ca_1,\ \cb_1	
\end{equation}
do not depend on $u_i,\,i\ge 0$.

Let us show that the conditions of Definition~\ref{pfa} hold, if we set 
$$
\CA_k=\CA'_k\cup\{1\},\quad 
\CB_k=\CB'_k\cup\{1\}. 
$$
Indeed, Condition~\ref{cmain} follows from Theorem~\ref{cpkn}.
Conditions~\ref{cm},~\ref{csym},~\ref{cir} hold because, by Lemma~\ref{repl}, 
Lemmas~\ref{lcm},~\ref{lsym},~\ref{lcir} are applicable to the canonical form 
of coverings described in Theorem~\ref{cpkn}.

Vector fields~\er{abpkn} satisfy~\er{ckkn} if and only if certain 
Lie algebra relations hold for~\er{abi}. 
Denote by $\mf^{KN}_{k-2}$ the quotient of the free Lie algebra 
generated by letters~\er{abi} over these relations. 
We obtain the system of fundamental algebras
\begin{equation}
\label{fkkn}
\dots\to\mf^{KN}_{n+1}\to\mf^{KN}_n\to\dots\to\mf^{KN}_1\to\mf^{KN}_0
\end{equation}
for equation~\er{kn}. 

\begin{proposition}[\cite{kncl}]
For each integer $n\ge 2$ there is a conserved current 
$D_tf_n=D_xg_n$ of the form 
\begin{equation*}
f_n=\frac{u_n^2}{u_1^2}+\tilde f_n(u,\dots,u_{n-1}),
\quad 
g_n=2\frac{u_nu_{n+2}}{u_1^2}+\tilde g_n(u,\dots,u_{n+1}),
\end{equation*}
where $\tilde f_n,\,\tilde g_n$ are polynomials 
in $1/u_1,\,u_i,\,i\ge 0$.
\end{proposition}
Similarly to Lemma~\ref{gcf}, we can find equivalent 
conserved currents
\begin{equation*}
	f'_n=f_n+D_x(h_n(u,\dots,u_{n-2})),\quad
	g'_n=g_n+D_t(h_n(u,\dots,u_{n-2}))
\end{equation*}
such that 
\begin{itemize}
	\item the functions $f'_n,\,g'_n$ are perfect,
	\item we have 
	\begin{equation*}
	\frac{\pd f'_n}{\pd u_s}(u,\dots,u_{s-1},0,\dots,0)=0\quad 
	\forall\,s\ge 2,
\end{equation*}
  \item $f'_n$ is polynomial in~\er{var1} with zero coefficient  
  at the monomial $u_1$.
\end{itemize}

\begin{example}
We have
\begin{gather*}
	f'_2=\frac{u_2^2}{u_1^2}+\frac23\frac{h(u)}{u_1^2},\\
	g'_2=2\frac{u_2u_4}{u_1^2}-\frac{u_3^2}{u_1^2}-\frac43\frac{h(u)u_3}{u_1^3}-4\frac{u_2^2u_3}{u_1^3}+\\
	\notag
	 +\frac94\frac{u_2^4}{u_1^4}-h(u)\frac{u_2^2}{u_1^4}+
2\frac{d h(u)}{d u}\frac{u_2}{u_1^2}-\frac13\frac{h(u)^2}{u_1^4}.
\end{gather*}
\end{example}


Return to algebras~\er{fkkn}. 
Let $\ma_k\subset\mf_{k-2}^{KN}$ be the subalgebra generated by 
$\ca_f,\,f\in\CA'_k$, and $\tilde\ma_k\subset\mf_{k-2}^{KN}$ be the subalgebra 
generated by $\ma_{k-2}$ and $\ca_1$. 
Similarly to Lemma~\ref{lbia}, we obtain
\begin{gather}
\label{biakn}
\cb_g\in\ma_{k-2}\quad\forall\,g\in\CB'_k,\\
\label{btaa}
[\cb_1,\tilde\ma_{k-2}]\subset\ma_{k-2}. 	
\end{gather} 

For $k\ge 5$ vector fields~\er{abpkn} can be rewritten as follows
\begin{gather}
\label{aspkn}
A=f'_{k-2}\ca^{k-2}+\frac{u_{k-3}^2}{u_1^2}\tilde\ca^{k-2}+A_0(u,\dots,u_{k-4}),\\	
\label{bspkn}
B=g'_{k-2}\ca^{k-2}+B_0(u,\dots,u_{k-1}),
\end{gather}
where 
\begin{gather}
\notag
\ca^{k-2}=\ca_{{u_{k-2}^2}{u_1^{-2}}},\quad 
\tilde\ca^{k-2}=\ca_{{u_{k-3}^2}{u_1^{-2}}},\\
\label{akakn}
[\ca^{k-2},A]=0,\\
\label{akbkn}
[\ca^{k-2},B]=3[A_0,\tilde\ca^{k-2}].	
\end{gather}

Equation~\er{akakn} implies
\begin{equation*}
	[\ca^{k-2},\tilde\ma_{k-2}]=0.
\end{equation*}
Combining this with~\er{biakn} and~\er{akbkn},
we obtain
\begin{gather}
\label{akb0kn}
[\ca^{k-2},\cb_1]=3[\ca_1,\tilde\ca^{k-2}],\\
\label{taa}
[\tilde\ca^{k-2},\ma_{k-2}]=0.
\end{gather}
Taking into account~\er{taa},~\er{btaa} and applying
$\ad^s\cb_1$ to~\er{akb0kn}, we obtain
\begin{equation}
\label{akbskn}
	-(\mathrm{ad}^{s+1}\cb_1)(\ca^{k-2})=
	3[\ca_1,(\mathrm{ad}^{s}\cb_1)(\tilde\ca^{k-2})]\quad
	\forall\,s\ge 0.
\end{equation}
Similarly to Section~\ref{cfakdv}, the obtained identities
imply that for each $k\ge 5$ the algebra $\mf_{k-2}^{KN}$ is obtained 
from $\mf_{k-3}^{KN}$ applying several times the operation 
of one-dimensional central extension.

Let us describe the algebras $\mf_{i}^{KN}$ for $i=0,1,2$. 
\begin{theorem}
\label{cgkn}
Any covering of equation~\eqref{kn}
of the form
\begin{gather*}
        D_xB-D_tA+[A,B]=0,\\
\label{dependkn}\notag
        A=A(u,u_1,u_2,u_3,u_4),\quad
        B=B(u,u_1,u_2,u_3,u_4)
\end{gather*}
is locally equivalent to a covering of the form
\begin{gather*}
A=f'_2C+\frac{1}{u_1}A_1(u)+V_1,\\
B=g'_2C-\frac{u_3}{u_1^2}A_1+ 
  \frac{u_2^2}{2u_1^3}A_1+
  \frac{2u_2}{u_1}\frac{\partial A_1}{\partial u}
  -\frac{h(u)}{3u_1^3}A_1+\\
  +\frac{2}{u_1}[A_1,\frac{\partial A_1}{\partial u}]
  -2u_1\frac{\partial^2 A_1}{\partial u^2}+V_2,
\end{gather*}
where $A_1=A_{10}+uA_{11}+u^2A_{12}$, the vector fields $C,\,V_i,\,A_{1k}$ 
do not depend on $u_i,\,i\ge 0$, and are subject to the following relations   
\begin{gather}
\label{crelkn}
        [C,V_i]=[C,A_{1k}]=[V_1,V_2]=[V_i,A_{1k}]=0\quad i=1,2,\quad k=0,1,2,\\
        \label{arel}
        2h(u)\frac{\partial A_1}{\partial u}-\frac{d h(u)}{d u}A_1-
        3[A_1,[A_1,\frac{\partial A_1}{\partial u}]]=0.
\end{gather}
\end{theorem}
\begin{proof}
This is proved by a straightforward computation following the scheme 
of the proof of Theorem~\ref{cpkn}. 
Relation~\er{arel} was obtained in~\cite{igonin}.	
\end{proof}

Equation~\er{arel} determines some relations between the vector fields $A_{1k},\,k=0,1,2$.
Let us describe the quotient of the free Lie algebra generated by
$A_{1k}$ over these relations.

Consider the ideal $\mathcal{I}\subset\mathbb{C}[v_1,v_2,v_3]$ generated by the
polynomials
\begin{equation}
  \label{elc}
  v_i^2-v_j^2+\frac83(e_j-e_i),\quad i,\,j=1,2,3,
\end{equation}
where $e_1,\,e_2,\,e_3$ are the roots of the polynomial $h(u)$.
Set
$$
E=\mathbb{C}[v_1,v_2,v_3]/\mathcal{I}.
$$
That is, $E$ is the ring of regular
functions on the affine elliptic curve in $\mathbb{C}^3$ defined
by polynomials~\eqref{elc}.
The image of $v_j\in\mathbb{C}[v_1,v_2,v_3]$
in $E$ is denoted by $\bar v_j$.
Consider also a basis $x_1,\,x_2,\,x_3$ of the Lie algebra
$\mathfrak{sl}_2(\mathbb{C})\cong\mathfrak{so}_3(\mathbb{C})$
with the relations
\begin{equation}
\label{xyz}
[x_1,x_2]=x_3,\quad [x_2,x_3]=x_1,\quad [x_3,x_1]=x_2
\end{equation}
and endow the space $L=\mathfrak{sl}_2\otimes_\mathbb{C} E$ with
the Lie algebra structure described in~\er{ga}.

\begin{proposition}[\cite{igonin}]
\label{R}
Suppose that the roots $e_1,\,e_2,\,e_3$ of $h(u)$ are distinct.
The quotient of the free Lie algebra generated by $A_{1k},\,k=0,1,2,$
over relations~\er{arel} is isomorphic to
the subalgebra $\mR\subset L$ generated by the elements
$$
x_1\otimes\bar v_1,\,x_2\otimes\bar v_2,\,x_3\otimes\bar v_3\in L.
$$
\end{proposition}

From~\er{crelkn} we obtain
\begin{equation*}
\mf_{0}^{KN}=0,\quad
\mf_{1}^{KN}\cong \mR\oplus\Com^2,\quad
\mf_{2}^{KN}\cong \mR\oplus\Com^3.
\end{equation*}

\begin{theorem}
\label{ftkn}
The algebra $\mR$ is quasi-finite.
\end{theorem}
\begin{proof}
Below we assume everywhere that $\{j,k,l\}=\{1,2,3\}$.
For each $j=1,2,3$ consider the subspace
$V_j\subset\mathbb{C}[v_1,v_2,v_3]$ spanned by
the monomials $v_j^{d_j}v_k^{d_k}v_l^{d_l}$ satisfying
\begin{equation}
\label{ddmod}
d_j\equiv d_k+1\equiv d_l+1\mod 2.
\end{equation}
Denote by $R_j$ the image of $V_j$ in the quotient space $E$.

The algebra $\mR$ was also studied in \cite{ll}
in connection with coverings of the Landau-Lifshitz equation.
In the proof of Lemma~3.1 of~\cite{ll} it is shown that
$\mR=\oplus_{j=1}^3\langle x_j\rangle\otimes R_j$.

Let $\mh\subset\mR$ be a subalgebra of finite codimension.
Then the subspace
$H_j=\{f\in R_j\,|\,x_j\otimes f\in\mh\}$ is of finite codimension
in $R_j$ for each $j=1,2,3$.
In addition,
from the definition of $R_j$ and relations~\eqref{xyz} we have
\begin{equation}
\label{hhh}
R_jR_k\subset R_l,\quad
H_jH_k\subset H_l.
\end{equation}
This implies that for all $j=1,2,3$ the subspace
\begin{equation}
\label{h'}
H'_j=\{a\in H_j\,|\,aR_k\subset H_l,\ aR_l\subset H_k\}
\end{equation}
is also of finite codimension in $R_j$.
From \eqref{hhh} and \eqref{h'} one gets
\begin{equation}
\label{hhh'}
H'_jH'_k\subset H'_l,\quad
H'_jR_jR_kR_l\subset H'_j.
\end{equation}

It is easy to see that $R_j=\langle \bar v_j\rangle+R_kR_l$.
Therefore,
\begin{equation}
\label{r2j}
    R^2_j=\langle\bar v^2_j\rangle+R_jR_kR_l.
\end{equation}
For each $j=1,2,3$ the subspace
\begin{equation}
\label{h''}
H''_j=\bigl\{a\in H'_j\,|\,a\bar v_k^2\subset H'_j,\ a\bar v_l^2\subset H'_j,\
aR_k\subset H'_l,\ aR_l\subset H'_k\bigl\}
\end{equation}
is of finite codimension in $H'_j$ and, therefore, in $R_j$.
By definitions~\eqref{h''}, \eqref{h'}
and properties~\eqref{hhh}, \eqref{hhh'}, \er{r2j},
one gets
\begin{equation*}
\label{ide}
R_kH''_j\subset H''_l,\quad R_lH''_j\subset H''_k,
\end{equation*}
which implies that
$\oplus_{j=1}^3\langle x_j\rangle\otimes H''_j\subset\mh$
is an ideal of $\mR$.
Since $H''_j$ is of finite codimension in $R_j$, this ideal
is of finite codimension in $\mR$.
\end{proof}

Collecting the results of this subsection and taking into account
Theorems~\ref{ftkn} and~\ref{qfce}, 
one obtains the following.
\begin{theorem}
\label{fakn}
The nonsingular Krichever-Novikov equation~\eqref{kn} possesses
fundamental algebras~\er{fkkn}, where $\mf^{KN}_0=0$.  
Each $\mf^{KN}_n$ for $n>0$ is quasi-finite and is obtained from $\mR$
applying several times the operation of one-dimensional 
central extension. 
\end{theorem}

\section{Coverings of the equation $u_t=u_{xxx}$}

In this section we study the linear equation 
\begin{equation}
\label{utu3}
u_t=u_{xxx}.
\end{equation}
The following theorem is proved by a straightforward computation. 
\begin{theorem}
\label{u3we}
Any Wahlquist-Estabrook covering 
\begin{gather*}
        D_xB-D_tA+[A,B]=0,\\
        A=A(u,u_1,u_2),\quad
        B=B(u,u_1,u_2)
\end{gather*}
of equation~\er{utu3} is of the form
\begin{gather*}
A=u^2A_2+uA_1+A_0,\\
B=u_2(2uA_2+A_1)-u_1^2A_2+u_1[A_1,A_0]-\frac12u^2[A_1,[A_1,A_0]]+\\
+u[A_0,[A_0,A_1]]+B_0,
\end{gather*}
where the vector fields $A_i,\,B_0$ depend only on $w^1,\dots,w^m$ and 
are subject to the relations 
\begin{gather}
        \label{a012}
        [A_0,A_2]=[A_1,A_2]=0,\\
\label{a0b0}
[A_0,B_0]=0,\\
\label{a1110}
[A_1,[A_1,[A_1,A_0]]]=0,\\
        \label{a0110}
[A_2,B_0]=\frac32[A_0,[A_1,[A_1,A_0]]],\\
\label{ab01}
[B_0,A_1]=[A_0,[A_0,[A_0,A_1]]].
\end{gather}
\end{theorem}
Denote by $\mN$ the quotient of the free Lie algebra generated by~$A_i,\,B_0$ 
over relations~\er{a012}, \er{a0b0}, \er{a1110}, \er{a0110}, \er{ab01}. 
Similarly to Section~\ref{cfkdv} one proves the following. 
\begin{theorem}
\label{fau3}
Equation~\er{utu3} possesses a system of fundamental algebras, which 
are obtained from $\mN$ applying several times the operation of one-dimensional 
central extension.
\end{theorem} 

Let us present some information on the structure of~$\mN$. 
\begin{theorem}
\label{ni}
There are ideals $\mN_i,\,i\in\zp$, of $\mN$ such that 
\begin{itemize}
\item $\mN_0=0,\quad \mN_i\subset\mN_{i+1}\quad \forall\,i\in\zp$,
\item the quotient $\mN_{i+1}/\mN_i$ is commutative for all $i\in\zp$,
\item the quotient $\mN/\cup_i\mN_i$ is solvable. 
\end{itemize}
\end{theorem}
\begin{proof}
For a subset $S$ of a Lie algebra we denote by $\langle S\rangle$ the ideal 
generated by this subset.
For simplicity, below the images of $A_i,\,B_0\in\mN$ in quotients 
of $\mN$ are denoted by the same symbols $A_i,\,B_0$. 
From the relations that define the algebra~$\mN$ one easily obtains 
the following. 
\begin{lemma} 
\label{commut}
Let $Q$ be a quotient algebra of $\mN$ and $C$ be an element of the subalgebra 
of $Q$ generated by $A_i$.
If $[A_0,C]=[A_1,C]=0$ then the ideal $\langle C\rangle \subset Q$ 
is spanned by the elements $(\ad^kB_0)(C),\,k\in\zp$, and  
is commutative. 
\end{lemma}
Let us construct the required ideals $\mN_i$. 
Set $\mN_0=0$ and $\mN_1=\langle A_2\rangle$. 
Combining relation~\er{a012} with the above lemma, we see that 
$\mN_1/\mN_0=\mN_1$ is commutative. 
By induction on $i\in\mathbb{N}$, set 
\begin{equation}
\label{Ni}
\mN_{i+1}=\langle \mN_i,[A_1,(\ad^{2i-1}A_0)(A_1)]\rangle.	
\end{equation}
\begin{lemma} 
For all $i\ge 1$ in the quotient algebra~$\mN/\mN_i$ we have 
\begin{gather}
\label{Ai}
[(\ad^kA_0)(A_1),(\ad^lA_0)(A_1)]=0\quad\forall\, k,l\in\zp\,\ k+l\le 2i-2,\\
\label{Ci}
[A_0,[A_1,(\ad^{2i-1}A_0)(A_1)]]=[A_1,[A_1,(\ad^{2i-1}A_0)(A_1)]]=0.
\end{gather}
\end{lemma}
\begin{proof}
Let us prove this by induction on $i$. For $i=1$ relation~\er{Ai} is trivial, and 
relation~\er{Ci} follows from~\er{a1110} and~\er{a0110}.
Suppose that the statement holds for $i=n\ge 1$ and let us prove it for $i=n+1$. 
By the induction assumption, relations~\er{Ai} for $i=n$ hold in~$\mN/\mN_{n+1}$. 
By definition~\er{Ni}, we have also
\begin{equation}
\label{Ni0}
[A_1,(\ad^{2n-1}A_0)(A_1)]=0. 
\end{equation}
Applying the Jacobi identity to~\er{Ni0} and taking into account~\er{Ai} for $i=n$, 
we obtain 
\begin{equation}
\label{Ai1}
[(\ad^kA_0)(A_1),(\ad^lA_0)(A_1)]=0\quad\forall\, k,l\in\zp\,\ k+l\le 2n-1.
\end{equation}
By the same argument, we have 
\begin{multline}
\label{Ai11}
[(\ad^kA_0)(A_1),(\ad^lA_0)(A_1)]=\\
=-[(\ad^{k+1}A_0)(A_1),(\ad^{l-1}A_0)(A_1)]\quad\forall\ k+l=2n.
\end{multline}
Using this, we obtain 
\begin{multline}
\label{Ai111}
[(\ad^kA_0)(A_1),(\ad^lA_0)(A_1)]=\\
=[(\ad^{l}A_0)(A_1),(\ad^{k}A_0)(A_1)]=0\quad\forall\ k+l=2n.
\end{multline}
Relations~\er{Ai1} and~\er{Ai111} imply~\er{Ai} for $i=n+1$. 

It remains to prove~\er{Ci} for $i=n+1$, that is,  
\begin{gather}
\label{Ci1}
[A_0,[A_1,(\ad^{2n+1}A_0)(A_1)]]=0,\\
\label{Ci2}
[A_1,[A_1,(\ad^{2n+1}A_0)(A_1)]]=0.
\end{gather}
Relation~\er{Ci2} follows easily from the Jacobi identity combined with~\er{Ai} for $i=n+1$. 

Similarly to~\er{Ai11} we have 
\begin{multline}
\label{Ai11i}
[(\ad^kA_0)(A_1),(\ad^lA_0)(A_1)]=\\
=-[(\ad^{k+1}A_0)(A_1),(\ad^{l-1}A_0)(A_1)]\quad\forall\ k+l=2n+1.
\end{multline}
Set $I=[A_1,(\ad^{2n+1}A_0)(A_1)]]$. Using~\er{Ai11i}, one gets 
\begin{multline*}
I=(-1)^{n}[(\ad^{n}A_0)(A_1),(\ad^{n+1}A_0)(A_1)]=\\
=(-1)^{n+1}[(\ad^{n-1}A_0)(A_1),(\ad^{n+2}A_0)(A_1)].
\end{multline*}
Applying $\ad A_0$ to this equality, we obtain 
\begin{multline}
\label{IAA}
[A_0,I]=(-1)^{n}[(\ad^{n}A_0)(A_1),(\ad^{n+2}A_0)(A_1)]=\\
=(-1)^{n+1}\bigl([(\ad^{n}A_0)(A_1),(\ad^{n+2}A_0)(A_1)]+\\
+[(\ad^{n-1}A_0)(A_1),(\ad^{n+3}A_0)(A_1)]\bigl).
\end{multline}

On the other hand, applying $\ad B_0$ to $[(\ad^{n-1}A_0)(A_1),(\ad^{n}A_0)(A_1)]=0$ 
and taking into account~\er{ab01} and~\er{a0b0}, one gets 
\begin{equation*}
[(\ad^{n+2}A_0)(A_1),(\ad^{n}A_0)(A_1)]+[(\ad^{n-1}A_0)(A_1),(\ad^{n+3}A_0)(A_1)]=0. 
\end{equation*}
Combining this with~\er{IAA}, we obtain $[A_0,I]=0$, which proves relation~\er{Ci2}. 
\end{proof}
By Lemma~\ref{commut}, relation~\er{Ci} implies that $\mN_{i+1}/\mN_i$ 
is commutative. Relation~\er{Ai} says that in the quotient algebra 
$\mN/\cup_i\mN_i$ we have 
\begin{equation}
\label{aaa0}
[(\ad^kA_0)(A_1),(\ad^lA_0)(A_1)]=0\quad\forall\, k,l\in\zp, 	
\end{equation}
which implies that this quotient of $\mN$ is solvable. 
\end{proof}
\begin{theorem} 
The algebra $\mN$ is not quasi-finite. 
\end{theorem} 
\begin{proof}
In the quotient algebra $\mN/\cup_i\mN_i$ denote $c_k=(\ad^kA_0)(A_1)$. 
Consider the subalgebra $\mg$ of $\mN/\cup_i\mN_i$ generated by 
$B_0$ and $c_k$. 
Obviously, for a quasi-finite algebra any quotient algebra 
and any subalgebra of finite codimension are also quasi-finite. 
Therefore, it is sufficient to prove that the algebra $\mg$ is not 
quasi-finite. 

Relations~\er{aaa0} say that $[c_k,c_l]=0$, while relations~\er{ab01} and~\er{a0b0} 
imply $[B_0,c_k]=c_{k+3}$. 
Let $m_k,\,k\in\zp$, be a sequence of nonzero complex numbers satisfying~$m_{k+3}=-(k+1)m_k$.  
Consider the following transitive action of $\mg$ 
on the manifold $M=\{(x,y)\in\Com^2\,|\,x\neq 0,\,y\neq 0\}$ 
\begin{equation*}
c_k\mapsto \frac{m_k}{x^{k+1}}\frac{\partial}{\partial y},
\quad B_0\mapsto\frac{1}{x^2}\frac{\partial}{\partial x}.
\end{equation*}
By Theorem~\ref{gih}, since the image of $\mg$ in $D(M)$ 
is infinite-dimensional, the algebra $\mg$ is not quasi-finite. 
\end{proof}

\section{Non-existence results for B\"acklund transformations}

\begin{theorem}
Equation~\er{utu3} is not connected by any B\"acklund transformation neither 
with the KdV equation nor with the nonsingular Krichever-Novikov equation.
\end{theorem}
\begin{proof} 
Below a Lie subalgebra denoted by $\mh$, $\mh^1$, or $\mh^2$ 
is always supposed to be of finite codimension. 
The following lemma is obvious. 
\begin{lemma}
\label{gggh}
Let $\mg$ be a finite-dimensional semisimple Lie algebra. 
Suppose that a Lie algebra $\mg_1$ is obtained from a Lie algebra 
$\mg_2$ applying several times the operation of one-dimensional central extension. 
Then each of the following properties holds for $i=1$ if and only if it holds for $i=2$. 
\begin{itemize}
\item There are a subalgebra $\mh\subset\mg_i$ and an epimorphism $\mh\to\mg$.  
\item For any subalgebra $\mh\subset\mg_i$ there is an epimorphism $\mh\to\mg$. 	
\end{itemize}
\end{lemma}
Set $\mg=\mathfrak{sl}_2(\mathbb{C})$. 
Let us prove first that there is no B\"acklund transformation between 
equation~\er{utu3} and the nonsingular Krichever-Novikov equation. 
Combining Lemma~\ref{gggh} with Theorems~\ref{fakn},~\ref{fau3}, and~\ref{cbt}, we see 
that it is sufficient to prove that for any subalgebras 
$\mh^1\subset\mR,\,\mh^2\subset\mN$ there is an epimorphism 
$\mh^1\to\mg$, but there is no epimorphism $\mh^2\to\mg$. 

There is a natural family of epimorphisms $\mR\to\mg$ 
parameterized by the points of the affine curve in $\mathbb{C}^3$ 
given by polynomials~\er{elc}. Namely, 
for a point $(a_1,a_2,a_3)$ of the curve the generator 
$x_i\otimes \bar v_i$ of $\mR$ is mapped to $a_ix_i\in\mg$. 
Since $\mh^1$ is of finite codimension in~$\mR$, there are polynomials  
$f_i(\bar v_1,\bar v_2,\bar v_3)$ and a point $(a_1,a_2,a_3)$ of the curve such that 
$x_i\otimes f_i$ belongs to $\mh^1$ and $f_i(a_1,a_2,a_3)\neq 0$ for all $i=1,2,3$. 
Then the restriction to $\mh^1$ of the corresponding homomorphism $\rho\colon\mR\to\mg$ 
is surjective, since the elements 
$$
\rho(x_i\otimes f_i)=f_i(a_1,a_2,a_3)x_i,\quad i=1,2,3, 
$$
span $\mg$. 
 
Non-existence of an epimorphism $\mh^2\to\mg$ follows 
from Theorem~\ref{ni}. Indeed, suppose that there is an epimorphism 
$\rho\colon\mh^2\to\mg$. Since $\mh^2\cap\mN_i$ is solvable, 
we have $\rho(\mh^2\cap\mN_i)=0$ for all $i$. Therefore, 
there is an epimorphism 
$$
\mh^2/\bigl(\mh^2\cap(\cup_i\mN_i)\bigl)\to\mg, 
$$
which is impossible, since $\mN/\cup_i\mN_i$ is solvable. 

Let us now prove that there is no  B\"acklund transformation between 
equation~\er{utu3} and the KdV equation. 
Since, according to Theorem~\ref{mtkdv}, each fundamental 
algebra of the KdV equation is obtained from $\mg\otimes\Com[\la]$ 
applying several times the operation of one-dimensional central extension, 
it is sufficient to prove that for any subalgebra $\mh^1\subset\mg\otimes\Com[\la]$ 
there is an epimorphism $\mh^1\to\mg$. Consider the natural family of epimorphisms 
$$
\rho_a\colon\mg\otimes\Com[\la]\to\mg,\quad 
g\otimes f(\la)\mapsto f(a)g,\quad a\in\Com.
$$
Since $\mh^1$ is of finite codimension, 
for some of these epimorphisms its restriction to $\mh^1$ is surjective.
\end{proof}

\section*{Acknowledgements}
The author is deeply grateful to I.~S.~Krasilshchik 
for careful reading of the paper and many valuable comments. 
The author thanks also R.~Martini and V.~V.~Sokolov  
for useful discussions.


\begin{thebibliography}{99}

\bibitem{rb}
A.~V. Bocharov, V.~N. Chetverikov, S.~V. Duzhin, N.~G. Khor{\cprime}kova,
I.~S. Krasil{\cprime}shchik, A.~V. Samokhin, Yu.~N. Torkhov, A.~M.
Verbovetsky, and  A.~M. Vinogradov.
\emph{Symmetries and Conservation Laws for Differential Equations of
Mathematical Physics}.
Amer. Math. Soc., Providence, RI, 1999.




\bibitem{dodd}
R.~Dodd and A.~Fordy.
The prolongation structures of quasipolynomial flows.
\emph{Proc. Roy. Soc. London Ser. A} \textbf{385} (1983), 389--429.




\bibitem{finley}
J.~D.~Finley and J.~K.~McIver. 
Prolongations to higher jets of Estabrook-Wahlquist coverings for PDEs. 
\emph{Acta Appl. Math.} \textbf{32} (1993), 197--225.

\bibitem{lie}
V.~V.~Gorbatsevich, A.~L.~Onishchik, and E.~B.~Vinberg.
\emph{Foundations of Lie theory and Lie transformation groups}.
Springer-Verlag, Berlin, 1997.

\bibitem{hc_zcr}
S.~Igonin.
Horizontal cohomology with coefficients 
and nonlinear zero-curvature representations.
\emph{Russian Math. Surveys} \textbf{58} (2003), 180--182

\bibitem{igonin}
S.~Igonin and R.~Martini.
Prolongation structure of the Krichever-Novikov equation.
\emph{J. Phys. A: Math. Gen.} \textbf{35} (2002), 9801-9810; 
{arxiv.org/nlin.SI/0208006}




\bibitem{kirn}
E.~G.~Kirnasov.
On Wahlquist-Estabrook type coverings over the heat equation.
\emph{Math. Notes} \textbf{42} (1987), 732--739.

\bibitem{nonl} I.~S.~Krasilshchik and A.~M.~Vinogradov. Nonlocal trends
in the geometry of differential equations. \emph{Acta Appl. Math.}
\textbf{15} (1989), 161-209.

\bibitem{86}
I.~S.~Krasil\cprime shchik, V.~V.~Lychagin, A.~M.~Vinogradov. 
\emph{Geometry of jet spaces and nonlinear partial differential 
equations.} Gordon \& Breach, New York, 1986. 


\bibitem{kv}
J.~Krasil\cprime shchik and A.~Verbovetsky.
Homological Methods in Equations of Mathematical Physics.
{arxiv.org/math.DG/9808130}


\bibitem{krich}
I.~M.~Krichever and  S.~P.~Novikov.
Holomorphic bundles over algebraic curves, and nonlinear equations.
\emph{Russian Math. Surveys} \textbf{35} (1980), 53--80.

\bibitem{marvan}
M.~Marvan. On zero-curvature representations of partial differential equations. 
\emph{Differential geometry and its applications (Opava, 1992)}, 103-122.  
\emph{Math. Publ.}, \textbf{1}, Silesian Univ. Opava, 1993.

\bibitem{orbits}
T.~Nagano. 
Linear differential systems with singularities 
and an application to transitive Lie algebras. 
\emph{J. Math. Soc. Japan} \textbf{18} (1966), 
398--404.




\bibitem{ll}
G.~H.~M.~Roelofs and R.~Martini.
Prolongation structure of the Landau-Lifshitz equation.
\emph{J. Math. Phys.} \textbf{34} (1993), 2394--2399.

\bibitem{backl_new}
C.~Rogers and W.~K.~Schief.
\emph{B\"acklund and Darboux transformations.} 
Cambridge Univ. Press, Cambridge, 2002.

\bibitem{backlund}
C.~Rogers and W.~F.~Shadwick.
\emph{B\"acklund transformations and their applications}. 
Academic Press, New York, 1982.


\bibitem{kncl}
V.~V.~Sokolov.
On the hamiltonian property of the Krichever-Novikov equation.
\emph{Soviet. Math. Dokl.} \textbf{30} (1984), 44-46.


\bibitem{sok2}
S.~I.~Svinolupov, V.~V.~Sokolov, and R.~I.~Yamilov.
B\"acklund transformations for integrable evolution equations.
\emph{Soviet Math. Dokl.} \textbf{28} (1983), 165--168.


\bibitem{kdv} H.~N.~van Eck.
The explicit form of the Lie algebra of Wahlquist and
Estabrook. A presentation problem.
\emph{Nederl. Akad. Wetensch. Indag. Math.} \textbf{45}
(1983), 149--164.

\bibitem{kdv1} H.~N.~van Eck.
A non-Archimedean approach to prolongation theory.
\emph{Lett. Math. Phys.} \textbf{12} (1986), 231--239.

\bibitem{verb}
A.~Verbovetsky.
Notes on the horizontal cohomology. 
\emph{Secondary calculus and cohomological physics
(Moscow, 1997)}, 211-231. 
\emph{Contemp. Math.}, \textbf{219},
Amer. Math. Soc., Providence, RI, 1998.

\bibitem{Prol} H.~D.~Wahlquist and F.~B.~Estabrook.
Prolongation structures
of nonlinear evolution equations. \emph{J. Math. Phys.}
\textbf{16} (1975), 1-7.

\end{thebibliography}
\end{document}